\documentclass[aps,nofootinbib,prd,preprintnumbers]{revtex4}
\pdfoutput=1
\usepackage[colorlinks=true,citecolor=darkred,urlcolor=darkred, pdfborder={0 0 0}]{hyperref}
\usepackage{amsfonts}
\usepackage{amsmath}
\usepackage[english]{babel}
\usepackage{graphicx}
\usepackage{epsfig}
\usepackage{bm}
\usepackage{longtable}
\usepackage{verbatim}
\usepackage{longtable}
\usepackage[utf8]{inputenc}
\usepackage{color}
\usepackage[normalem]{ulem}
\definecolor{darkred}{rgb}{0.6,0,0}

\def \nbb {$\beta\beta_{0\nu}$ }
\def\vev#1{\left\langle #1\right\rangle}

\definecolor{greenLinks}{rgb}{0, 0.6, 0} 
\definecolor{blueLinks}{rgb}{0, 0, 0.6}
\definecolor{redLinks}{rgb}{0.6, 0, 0}
\definecolor{tempText}{rgb}{0.55, 0.10,0.67}
\definecolor{eprintLinks}{rgb}{0.4, 0.4, 0.4}
\definecolor{journalLinks}{rgb}{0.6, 0, 0}
\newcommand {\ignore}[1]{}

\newcommand{\jv}[1]{{\color{blue}  #1}}

\def\SM{$\mathrm{ SU(3)_C \otimes SU(2)_L \otimes U(1)_Y }$ }
\newcommand{\sm}{{Standard Model }}
\def\vev#1{\left\langle #1\right\rangle}

\def\lsim{\mathrel{\rlap{\lower4pt\hbox{\hskip1pt$\sim$}}
    \raise1pt\hbox{$<$}}}
\def\gsim{\mathrel{\rlap{\lower4pt\hbox{\hskip1pt$\sim$}}
    \raise1pt\hbox{$>$}}}
\def\LR{$\mathrm{SU(3) \otimes SU(2)_L \otimes SU(2)_R \otimes U(1)_{B-L}}$ }

\def\U1s{$\mathrm{U_{1}^{(a)}\otimes U_{1}^{(b)}\otimes U_{1}^{(c)}\otimes U_{1}^{(d)}\otimes U_{1}^{(e)}}$ }
\def\3211{$\mathrm{SU(3) \otimes SU(2)_L \otimes U(1)_R \otimes U(1)_{B-L}}$ }
\def\321{$\mathrm{SU(3) \otimes SU(2) \otimes U(1)}$ }
\def\422{$\mathrm{SU(4) \otimes SU(2) \otimes SU(2)_R}$ }

\def \nbb {$\rm 0\nu \beta \beta $ }

\newcommand{\mathsym}[1]{{}}

\input{tcilatex}

\begin{document}


\bibliographystyle{unsrt}
\title{Neutrino predictions from a left-right symmetric\\ flavored extension of the standard model}
\author{A. E. C\'{a}rcamo Hern\'{a}ndez${}^{a}$}
\email{antonio.carcamo@usm.cl}
\author{Sergey Kovalenko}
\email{sergey.kovalenko@usm.cl}
\author{Jos\'{e} W. F. Valle$^{{b}}$}
\email{jose.valle@ific.uv.es}
\author{C.A. Vaquera-Araujo$^{{c}}$,$^{{d}}$}
\email{vaquera@fisica.ugto.mx}
\affiliation{$^{{a}}$Universidad T\'{e}cnica Federico Santa Mar\'{\i}a\\
and Centro Cient\'{\i}fico-Tecnol\'{o}gico de Valpara\'{\i}so\\
Casilla 110-V, Valpara\'{\i}so, Chile,\\
$^{{b}}$AHEP Group, Instituto de F\'{\i}sica Corpuscular-CSIC/Universitat de Valencia, Parc Cientific de Paterna,\\
C/Catedr\'{a}tico Jos\'{e} Beltr\'{a}n, 2 E-46980 Paterna (Valencia) - Spain,\\
$^{{c}}$Departamento de F\'isica, DCI, Campus Le\'on, Universidad de
Guanajuato, Loma del Bosque 103, Lomas del Campestre C.P. 37150, Le\'on, Guanajuato, M\'exico,\\
$^{{d}}$Consejo Nacional de Ciencia y Tecnolog\'ia, Av. Insurgentes Sur 1582. Colonia Cr\'edito Constructor, Del. Benito Ju\'arez, C.P. 03940, Ciudad de M\'exico, M\'exico}
\date{\today }

\begin{abstract}
We propose a left-right symmetric electroweak extension of the \sm based on the $\Delta \left( 27\right)$ family symmetry. 
The masses of all electrically charged \sm fermions lighter than the top quark are induced by a Universal Seesaw mechanism mediated by exotic fermions. 
The top quark is the only \sm fermion to get mass directly from a tree level renormalizable Yukawa interaction, while neutrinos are unique in that they get calculable radiative masses through a low-scale seesaw mechanism.
The scheme has generalized $\mu-\tau$ symmetry and leads to a restricted range of neutrino oscillations parameters, with a nonzero neutrinoless double beta decay amplitude lying at the upper ranges generically associated to normal and inverted neutrino mass ordering.

\end{abstract}

\maketitle

\section{Introduction}
\label{sec:introduction}

The \SM gauge theory provides a remarkable description of the interactions of quarks and leptons as mediated by intermediate vector bosons associated to the \sm gauge structure. However, it is well-known to suffer from a number of drawbacks. Most noticeably, it fails to account for neutrino masses, needed to describe the current oscillation data\cite{deSalas:2017kay}. Likewise, it does not provide a dynamical understanding of the origin of parity violation in the weak interaction. Last, but not least, it also fails in providing an understanding of charged lepton and quark mass hierarchies and mixing angles from first principles. Left-right symmetric electroweak extensions of the Weinberg-Salam theory address the origin of parity violation~\cite{Pati:1974yy,Mohapatra:1974gc}, while models based on non-Abelian flavor symmetries~\cite{Ishimori:2010au} address the flavor issues~\cite{Morisi:2012fg,King:2014nza}. Combining these features is a desirable step forward. Indeed, a predictive Pati-Salam theory of fermion masses and mixing combining both approaches has been suggested recently~\cite{CarcamoHernandez:2017owh}.

In this paper we propose a less restrictive left-right symmetric electroweak extension of the \sm based on the \LR gauge group and the $\Delta \left( 27\right) $ family symmetry. 
Of the \sm fermions only the top quark acquires mass through a tree level renormalizable Yukawa interaction. Exotic charged fermions acquire mass from their corresponding tree level mass terms, while gauge singlet fermions can also have gauge-invariant tree level Majorana mass terms.
The masses for the other electrically charged \sm fermions, namely quarks lighter than the top, as well as charged leptons, are all induced by a Universal Seesaw mechanism mediated by the exotic fermions. 
The mass hierarchies as well as the quark mixing angles arise from the spontaneous breaking of the $\Delta \left( 27\right)\otimes Z_{6}\otimes Z_{12}$ discrete family group, and the radiative nature of the inverse seesaw mechanism is guaranteed by spontaneously broken $Z_{4}$ and $Z_{12}$ symmetries, with $Z_{12}$ spontaneously broken down to a preserved $Z_2$ symmetry.
The Cabibbo mixing arises from the up-type quark sector, whereas the down-type quark sector contributes to the remaining CKM mixing angles. On the other hand, the lepton mixing matrix receives its main contributions from the light active neutrino mass matrix, while the \sm charged lepton mass matrix provides Cabibbo-like corrections to these parameters. These features of our flavoured left-right symmetric theory regarding the origin of the main contributions to the fermionic mixing parameters
result from the discrete symmetries and hold in the particular basis of the eigenstates of the generators of the flavour group. 
Finally, the masses for the light active neutrinos emerge from a low-scale inverse/linear seesaw mechanism~\cite{Mohapatra:1986bd,GonzalezGarcia:1988rw,Akhmedov:1995vm,Akhmedov:1995ip,Malinsky:2005bi} with one-loop-induced seed mass parameters~\cite{Bazzocchi:2009kc,CarcamoHernandez:2017kra}.

\section{The model}
\label{model}

The model is based on the left-right gauge symmetry \LR supplemented by the discrete $\Delta \left( 27\right) \otimes Z_{4}\otimes Z_{6}\otimes Z_{12}$ family group. The particle content and gauge quantum numbers are summarized in table~\ref{tab:gqn1}, while the transformation properties of the fields under the discrete symmetries are presented in tables~\ref{tab:gqn2},~\ref{tab:gqn3} and ~\ref{tab:gqn4}. Here $\omega =e^{\frac{2\pi i}{3}}$ and the numbers in boldface denote the $\Delta \left( 27\right)$ irreducible representations.

Notice that the fermion sector of the original left-right symmetric model has been extended with two vectorlike up-type quarks $T_{k}$, $k=1,2$, three vectorlike down type quarks $B_{i}$, three vectorlike charged leptons $E_{i}$ and six neutral Majorana singlets $S_i$, $\Omega_i$, with $i=1,2,3$.
The role of the new exotic vectorlike fermions is to generate the masses for \sm charged fermions from a Universal Seesaw mechanism 
\jv{\cite{Davidson:1987mh,Berezhiani:1991ds,Sogami:1991yq,Gu:2010zv,Alvarado:2012xi,Hernandez:2013mcf,Kawasaki:2013apa,Mohapatra:2014qva,Dev:2015vjd,Borah:2017inr,Patra:2017gak,King:2018fcg,Babu:2018vrl}}. 
Neutrino masses are in turn produced by an inverse seesaw mechanism, triggered by a one loop generated mass scale~\cite{Bazzocchi:2009kc,CarcamoHernandez:2017kra} from the interplay of the gauge singlet fermions $S_i$ and $\Omega_i$.

In the scalar sector the model includes a bi-doublet, two $SU(2)_L$ doublets, and two $SU(2)_R$ doublets with vacuum expectation values (VEVs):
\begin{equation}
\vev{ \Phi } =\left( 
\begin{array}{cc}
v_{1} & 0 \\ 
0 & v_{2}%
\end{array}%
\right) ,\hspace{1.5cm}\vev{ \chi _{kL}} =\left( 
\begin{array}{c}
0 \\ 
v_{kL}%
\end{array}%
\right) ,\hspace{1.5cm}\vev{ \chi _{kR}} =\left( 
\begin{array}{c}
0 \\ 
v_{kR}%
\end{array}%
\right),\hspace{1.5cm}k=1,2,
\end{equation}%
as well as several singlet scalars 
\begin{equation}
\sigma,\, \eta,\, \varphi,\, \rho_{i},\, \phi _{i},\, \tau _{i},\, \xi
_{i}\,\hspace{1.5cm}i=1,2,3.
\end{equation}
In the following, we set $v_{2}=0$ for simplicity. We assume that all singlet scalar fields acquire nonvanishing vacuum expectation values, except for $\varphi$. 

\begin{table}[ht]
\centering
\begin{tabular}{|c||c|c|c|c||c|c|c|c|c|c|c|c||c|c|c||c|c|c|c|c|c|c|}
\hline\hline
Field & $Q_{iL}$ & $Q_{iR}$ & $L_{iL}$ & $L_{iR}$ & $T_{kL}$ & $T_{kR}$ & $%
B_{iL}$ & $B_{iR}$ & $E_{iL}$ & $E_{iR}$ & $S_{i}$ & $\Omega_{i}$ & $\Phi$ & 
$\chi_{kL}$ & $\chi_{kR}$ & $\sigma$ & $\eta$ & $\varphi$ & $\rho_{i}$ & $%
\phi _{i}$ & $\tau _{i}$ & $\xi _{i}$ \\ \hline
$SU\left( 3\right)_{c}$ & $\mathbf{3}$ & $\mathbf{3}$ & $\mathbf{1}$ & $%
\mathbf{1}$ & $\mathbf{3}$ & $\mathbf{3}$ & $\mathbf{3}$ & $\mathbf{3}$ & $%
\mathbf{1}$ & $\mathbf{1}$ & $\mathbf{1}$ & $\mathbf{1}$ & $\mathbf{1}$ & $%
\mathbf{1}$ & $\mathbf{1}$ & $\mathbf{1}$ & $\mathbf{1}$ & $\mathbf{1}$ & $%
\mathbf{1}$ & $\mathbf{1}$ & $\mathbf{1}$ & $\mathbf{1}$ \\ \hline
$SU\left( 2\right)_{L}$ & $\mathbf{2}$ & $\mathbf{1}$ & $\mathbf{2}$ & $%
\mathbf{1}$ & $\mathbf{1}$ & $\mathbf{1}$ & $\mathbf{1}$ & $\mathbf{1}$ & $%
\mathbf{1}$ & $\mathbf{1}$ & $\mathbf{1}$ & $\mathbf{1}$ & $\mathbf{2}$ & $%
\mathbf{2}$ & $\mathbf{1}$ & $\mathbf{1}$ & $\mathbf{1}$ & $\mathbf{1}$ & $%
\mathbf{1}$ & $\mathbf{1}$ & $\mathbf{1}$ & $\mathbf{1}$ \\ \hline
$SU\left( 2\right)_{R}$ & $\mathbf{1}$ & $\mathbf{2}$ & $\mathbf{1}$ & $%
\mathbf{2}$ & $\mathbf{1}$ & $\mathbf{1}$ & $\mathbf{1}$ & $\mathbf{1}$ & $%
\mathbf{1}$ & $\mathbf{1}$ & $\mathbf{1}$ & $\mathbf{1}$ & $\mathbf{2}$ & $%
\mathbf{1}$ & $\mathbf{2}$ & $\mathbf{1}$ & $\mathbf{1}$ & $\mathbf{1}$ & $%
\mathbf{1}$ & $\mathbf{1}$ & $\mathbf{1}$ & $\mathbf{1}$ \\ \hline
$U(1)_{B-L}$ & $\frac{1}{3}$ & $\frac{1}{3}$ & $-1$ & $-1$ & $\frac{4}{3}$ & 
$\frac{4}{3}$ & $-\frac{2}{3}$ & $-\frac{2}{3}$ & $-2$ & $-2$ & $0$ & $0$ & $%
0$ & $1$ & $1$ & $0$ & $0$ & $0$ & $0$ & $0$ & $0$ & $0$ \\ \hline\hline
\end{tabular}%
\caption{Particle content and transformation properties under the \LR gauge symmetries. Here $i=1,2,3$ and $k=1,2$.}
\label{tab:gqn1}
\end{table}
\begin{table}[ht]
\centering
\begin{tabular}{|c||c|c|c|c|c|c|c|c|c|c|c|c|c|c|c|c|}
\hline\hline
Field & $Q_{1L}$ & $Q_{1R}$ & $Q_{2L}$ & $Q_{2R}$ & $Q_{3L}$ & $Q_{3R}$ & $T_{1L}$ & $T_{1R}$ & $T_{2L}$ & $T_{2R}$ & $B_{1L}$ & $B_{1R}$ & $B_{2L}$ & 
$B_{2R}$ & $B_{3L}$ & $B_{3R}$  \\ \hline
$\Delta(27)$ & $\mathbf{1}_{\mathbf{0,0}}$ & $\mathbf{1}_{\mathbf{0,0}}$ & $\mathbf{1}_{\mathbf{0,0}}$ & $\mathbf{1}_{\mathbf{0,0}}$ & $\mathbf{1}_{\mathbf{2,1}}$ & $\mathbf{1}_{\mathbf{2,2}}$ & $\mathbf{1}_{\mathbf{0,0}}$ & $\mathbf{1}_{\mathbf{0,0}}$ & $\mathbf{1}_{\mathbf{0,0}}$ & $\mathbf{1}_{\mathbf{0,0}}$ & $\mathbf{1}_{\mathbf{0,0}}$ & $\mathbf{1}_{\mathbf{0,0}}$ & $\mathbf{1}_{\mathbf{0,0}}$ & $\mathbf{1}_{\mathbf{0,0}}$ & $\mathbf{1}_{\mathbf{0,0}}$ & $\mathbf{1}_{\mathbf{0,0}}$  \\ \hline
$Z_4$ & $1$ & $1$ & $1$ & $1$ & $1$ & $1$ & $1$ & $1$ & $1$ & $1$ & $1$ & $1$ & $1$ & $1$ & $1$ & $1$  \\ \hline
$Z_6$ & $-1$ & $-1$ & $\omega^2$ & $\omega^2$ & $1$ & $1$ & $\omega^{-\frac{1}{2}}$ & $\omega^{-\frac{1}{2}}$ & $\omega^2$ & $\omega^2$ & $\omega^{-\frac{1}{2}}$ & $\omega^{-\frac{1}{2}}$ & $\omega$ & $\omega$ & $1$ & $1$  \\ \hline
$Z_{12}$ & $1$ & $1$ & $1$ & $1$ & $1$ & $1$ & $-1$ & $-1$ & $1$ & $1$ & $-1$ & $-1$ & $i$ & $i$ & $1$ & $1$  \\ \hline
\end{tabular}%
\caption{Transformation properties of the quarks under the flavor symmetry $\Delta\left( 27\right) \otimes Z_{4}\otimes Z_{6}\otimes Z_{12}$.}
\label{tab:gqn2}
\end{table}
\begin{table}[ht]
\centering
\begin{tabular}{|c||c|c|c|c|c|c|c|c|c|c|c|c|c|c|}
\hline\hline
Field & $L_{1L}$ & $L_{1R}$ & $L_{2L}$ & $L_{2R}$ & $L_{3L}$ & $L_{3R}$ & $E_{1L}$ & $E_{1R}$ & $E_{2L}$ & $E_{2R}$ & $E_{3L}$ & $E_{3R}$ & $S$ & $\Omega$  \\ \hline
$\Delta(27)$ & $\mathbf{1}_{\mathbf{2,0}}$ & $\mathbf{1}_{\mathbf{2,0}}$ & $\mathbf{1}_{\mathbf{2,0}}$ & 
$\mathbf{1}_{\mathbf{2,0}}$ & $\mathbf{1}_{\mathbf{2,0}}$ & $\mathbf{1}_{\mathbf{2,0}}$ & $\mathbf{1}_{\mathbf{2,0}}$ & $\mathbf{1}_{\mathbf{2,0}}$ & $\mathbf{1}_{\mathbf{2,0}}$ & $\mathbf{1}_{\mathbf{2,0}}$ & $\mathbf{1}_{\mathbf{2,0}}$ & $\mathbf{1}_{\mathbf{2,0}}$ & $\mathbf{3}$ & $\overline{\mathbf{3}}$  \\ \hline
$Z_4$ & $1$ & $1$ & $1$ & $1$ & $1$ & $1$ & $1$ & $1$ & $1$ & $1$ & $1$ & $1$ & $1$ & $i$  \\ \hline
$Z_6$ & $-1$ & $-1$ & $\omega^2$ & $\omega^2$ & $\omega^{-\frac{1}{2}}$ & $\omega^{-\frac{1}{2}}$ & $\omega^{-\frac{1}{2}}$ & $\omega^{-\frac{1}{2}}$ & $\omega$ & $\omega$ & $\omega^{-\frac{1}{2}}$ & $\omega^{-\frac{1}{2}}$ & $1$ & $\omega^{-\frac{1}{2}}$  \\ \hline
$Z_{12}$ & $-1$ & $-1$ & $-1$ & $-1$ & $-1$ & $-1$ & $1$ & $1$ & $-i$ & $-i$ & $-1$ & $-1$ & $-\omega^{\frac{1}{2}}$ & $\omega^{-\frac{1}{2}}$  \\ \hline
\end{tabular}%
\caption{Transformation properties of the leptons under the flavor symmetry $\Delta\left( 27\right) \otimes Z_{4}\otimes Z_{6}\otimes Z_{12}$.}
\label{tab:gqn3}
\end{table}
\begin{table}[ht]
\centering
\begin{tabular}{|c||c|c|c|c|c|c|c|c|c|c|c|c|}
\hline\hline
Field & $\Phi$ & $\chi_{1L}$ & $\chi_{1R}$ & $\chi_{2L}$ & $\chi_{2R}$ & $\sigma$ & $\eta$ & $\varphi$ & $\rho$ & $\phi$ & $\tau$ & $\xi$  \\ \hline
$\Delta(27)$ & $\mathbf{1}_{\mathbf{0,2}}$ & $\mathbf{1}_{\mathbf{0,0}}$ & $\mathbf{1}_{\mathbf{0,0}}$ & 
 $\mathbf{1}_{\mathbf{2,1}}$ & $\mathbf{1}_{\mathbf{2,2}}$ & $\mathbf{1}_{\mathbf{0,0}}$ & $\mathbf{1}_{\mathbf{0,0}}$ & $\mathbf{1}_{\mathbf{0,0}}$ & $\overline{\mathbf{3}}$ & $\overline{\mathbf{3}}$ & $\overline{\mathbf{3}}$ & $\mathbf{3}$  \\ \hline
$Z_4$ & $1$ & $1$ & $1$ & $1$ & $1$ & $1$ & $1$ & $i$ & $1$ & $1$ & $1$ & $-1$  \\ \hline
$Z_6$ & $1$ & $1$ & $1$ & $1$ & $1$ & $\omega^{-\frac{1}{2}}$ & $\omega$ & $\omega^{-\frac{1}{2}}$ & $\omega$ & $-1$ & $1$ & $\omega^{2}$ \\ \hline
$Z_{12}$ & $1$ & $1$ & $1$ & $1$ & $1$ & $1$ & $-i$ & $-1$ & $\omega^{-\frac{1}{2}}$ & $\omega^{-\frac{1}{2}}$ & $\omega^{-\frac{1}{2}}$ & $\omega^{2}$  \\ \hline
\end{tabular}%
\caption{Transformation properties of the scalars under the flavor symmetry $\Delta\left( 27\right) \otimes Z_{4}\otimes Z_{6}\otimes Z_{12}$.}
\label{tab:gqn4}
\end{table}
\begin{table}[ht]
\centering
\begin{tabular}{|c||c|c||c|c|c||c|c|c|c|c|c|c|}
\hline\hline
Field &  $S_{i}$ & $\Omega_{i}$ & $\Phi$ & 
$\chi_{kL}$ & $\chi_{kR}$ & $\sigma$ & $\eta$ & $\varphi$ & $\rho_{i}$ & $%
\phi _{i}$ & $\tau _{i}$ & $\xi _{i}$ \\ \hline
$U_{1}^{(a)}$ & $0$ & $1$ & $0$ & $0$ & $0$ & $0$ & $0$ & $1$ & $0$ & $0$ & $0$ & $2$ \\ \hline
$U_{1}^{(b)}$  & $1$ & $0$ & $0$ & $0$ & $0$ & $0$ & $0$ & $1$ & $-1$ & $-1$ & $-1$ & $0$ \\ \hline
$U_{1}^{(c)}$ & $1$ & $0$ & $0$ & $1$ & $-1$ & $0$ & $0$ & $1$ & $0$ & $0$ & $0$ & $0$ \\ \hline
$U_{1}^{(d)}$ & $1$ & $1$ & $0$ & $0$ & $0$ & $0$ & $0$ & $2$ & $-1$ & $-1$ & $-1$ & $2$ \\ \hline
$U_{1}^{(e)}$ & $1$ & $1$ & $0$ & $1$ & $-1$ & $0$ & $0$ & $2$ & $0$ & $0$ & $0$ & $2$ \\ \hline\hline
\end{tabular}%
\caption{Accidental \U1s symmetries.}
\label{tab:acc}
\end{table}
Given the particle content, the following up-type, down-type quark, charged lepton and neutrino Yukawa terms arise, respectively:
\begin{eqnarray}
-\mathcal{L}_{Y}^{\left( U\right) } &=&\alpha \overline{Q}_{3L}\Phi
Q_{3R}+x_{11}\overline{Q}_{1L}\widetilde{\chi }_{1L}T_{1R}\frac{\eta ^{2}}{%
\Lambda ^{2}}+x_{22}\overline{Q}_{2L}\widetilde{\chi }_{1L}T_{2R}+x_{12}%
\overline{Q}_{1L}\widetilde{\chi }_{1L}T_{2R}\frac{\sigma }{\Lambda }  \notag
\\
&&+x_{11}\overline{T}_{1L}\widetilde{\chi }_{1R}^{\dagger }Q_{1R}\frac{%
\left( \eta ^{\ast }\right) ^{2}}{\Lambda ^{2}}+x_{22}\overline{T}_{2L}%
\widetilde{\chi }_{1R}^{\dagger }Q_{2R}+x_{12}\overline{T}_{2L}\widetilde{%
\chi }_{1R}^{\dagger }Q_{1R}\frac{\sigma ^{\ast }}{\Lambda }%
+\dsum\limits_{i=1}^{2}m_{T_{i}}\overline{T}_{iL}T_{iR}+h.c.,
\label{LYU}
\end{eqnarray}
\begin{eqnarray}
-\mathcal{L}_{Y}^{\left( D\right) } &=&y_{11}\overline{Q}_{1L}\chi
_{1L}B_{1R}\frac{\eta ^{2}}{\Lambda ^{2}}+y_{22}\overline{Q}_{2L}\chi
_{1L}B_{2R}\frac{\eta }{\Lambda }+y_{33}\overline{Q}_{3L}\chi _{2L}B_{3R} 
\notag \\
&&+y_{13}\overline{Q}_{1L}\chi _{1L}B_{3R}\frac{\sigma ^{3}}{\Lambda ^{3}}%
+y_{23}\overline{Q}_{2L}\chi _{1L}B_{3R}\frac{\sigma ^{2}}{\Lambda ^{2}} 
\notag \\
&&+y_{11}\overline{B}_{1L}\chi _{1R}^{\dagger }Q_{1R}\frac{\left( \eta
^{\ast }\right) ^{2}}{\Lambda ^{2}}+y_{22}\overline{B}_{2L}\chi
_{1R}^{\dagger }Q_{2R}\frac{\eta ^{\ast }}{\Lambda }+y_{33}\overline{B}%
_{3L}\chi _{2R}^{\dagger }Q_{3R}  \notag \\
&&+y_{13}^{\ast }\overline{B}_{3L}\chi _{1R}^{\dagger }Q_{1R}\frac{\left(
\sigma ^{\ast }\right) ^{3}}{\Lambda ^{3}}+y_{23}\overline{B}_{3L}\chi
_{1R}^{\dagger }Q_{2R}\frac{\left( \sigma ^{\ast }\right) ^{2}}{\Lambda ^{2}}%
+\dsum\limits_{i=1}^{3}m_{B_{i}}\overline{B}_{iL}B_{iR}+h.c.,
\label{LYD}
\end{eqnarray}
\begin{eqnarray}
-\mathcal{L}_{Y}^{\left( E\right) } &=&z_{11}\overline{L}_{1L}\chi
_{1L}E_{1R}\frac{\eta ^{2}}{\Lambda ^{2}}+z_{22}\overline{L}_{2L}\chi
_{1L}E_{2R}\frac{\eta }{\Lambda }+z_{33}\overline{L}_{3L}\chi _{1L}E_{3R} 
\notag \\
&&+z_{13}\overline{L}_{1L}\chi _{1L}E_{3R}\frac{\sigma
^{2}}{\Lambda ^{2}}+z_{23}\overline{L}_{2L}\chi _{1L}E_{3R}\frac{\sigma
}{\Lambda }  \notag \\
&&+z_{11}\overline{E}_{1L}\chi _{1R}^{\dagger }L_{1R}\frac{\left( \eta
^{\ast }\right) ^{2}}{\Lambda ^{2}}+z_{22}\overline{E}_{2L}\chi
_{1R}^{\dagger }L_{2R}\frac{\eta ^{\ast }}{\Lambda }+z_{33}\overline{E}%
_{3L}\chi _{1R}^{\dagger }L_{3R}  \notag \\
&&+z_{13}^{\ast }\overline{E}_{3L}\chi _{1R}^{\dagger
}L_{1R}\frac{\left( \sigma ^{\ast }\right) ^{2}}{\Lambda
^{2}}+z_{23}\overline{E}_{3L}\chi _{1R}^{\dagger }L_{2R}\frac{\sigma ^{\ast
}}{\Lambda }\color{black}{+}\dsum\limits_{i=1}^{3}m_{E_{i}}\overline{E}%
_{iL}E_{iR}+h.c.,
\label{LYE}
\end{eqnarray}
\begin{eqnarray}
-\mathcal{L}_{Y}^{\left( \nu \right) } &=&\gamma _{1}\overline{L}_{1L}\Phi
L_{1R}\frac{\left( \xi \xi ^{\ast }\right) _{\mathbf{1}_{\mathbf{0,1}}}}{%
\Lambda ^{2}}+\gamma _{2}\overline{L}_{2L}\Phi L_{2R}\frac{\left( \xi \xi
^{\ast }\right) _{\mathbf{1}_{\mathbf{0,1}}}}{\Lambda ^{2}}+\gamma _{3}%
\overline{L}_{3L}\Phi L_{3R}\frac{\left( \xi \xi ^{\ast }\right) _{\mathbf{1}%
_{\mathbf{0,1}}}}{\Lambda ^{2}}  \notag \\
&&+\frac{\kappa _{1}}{\Lambda }\overline{L}_{1L}\widetilde{\chi }_{1L}\left(
\rho S\right) _{\mathbf{1}_{\mathbf{2,0}}}\frac{\sigma ^{\ast }}{\Lambda }+%
\frac{\kappa _{2}}{\Lambda }\overline{L}_{2L}\widetilde{\chi }_{1L}\left(
\phi S\right) _{\mathbf{1}_{2\mathbf{,0}}}\frac{\sigma ^{\ast }}{\Lambda }+%
\frac{\kappa _{3}}{\Lambda }\overline{L}_{3L}\widetilde{\chi
}_{1L}\left( \tau S\right) _{\mathbf{1}_{2\mathbf{,0}}}\frac{\sigma
}{\Lambda }  \notag \\
&&+\frac{\kappa _{1}}{\Lambda }\left( \overline{S^{C}}\rho \right) _{\mathbf{%
1}_{1\mathbf{,0}}}\widetilde{\chi }_{1R}^{\dagger }L_{1R}\frac{\sigma ^{\ast
}}{\Lambda }+\frac{\kappa _{2}}{\Lambda }\left( \overline{S^{C}}\phi \right)
_{\mathbf{1}_{1\mathbf{,0}}}\widetilde{\chi }_{1R}^{\dagger }L_{2R}\frac{%
\sigma }{\Lambda }+\frac{\kappa _{3}}{\Lambda }\left(
\overline{S^{C}}\tau \right) _{\mathbf{1}_{1\mathbf{,0}}}\widetilde{\chi
}_{1R}^{\dagger }L_{3R}\frac{\sigma ^{\ast }}{\Lambda }  \notag \\
&&+\lambda _{1}\left( \overline{\Omega }\Omega ^{C}\right) _{%
\overline{\mathbf{3}}_{S_{1}}}\xi +\lambda _{2}\left( \overline{\Omega }%
\Omega ^{C}\right) _{\overline{\mathbf{3}}_{S_{2}}}\xi +\lambda _{3}\left( 
\overline{S}\Omega ^{C}\right) _{\mathbf{1}_{\mathbf{0,0}}}\varphi +
h.c.,
\label{LYnu}
\end{eqnarray}
Let us note that the neutrino Yukawa terms given in Eq.~(\ref{LYnu}) have accidental \U1s symmetries described in Table \ref{tab:acc}. These are spontaneously broken by the VEVs of the scalar fields charged under these symmetries. As a result, massless Goldstone bosons are expected to arise from the spontaneous breakdown of these symmetries. However, these can acquire masses from scalar interactions like $\lambda\rho^2\xi^2$ and $M\chi^{\dagger}_{2L}\Phi\chi_{2R}$, invariant under the symmetry group $\mathcal{G}$ of our model, but not under the accidental $U_{1}^{(a)}\times U_{1}^{(b)}\times U_{1}^{(d)}$ and $U_{1}^{(c)}\times U_{1}^{(e)}$ symmetries, respectively.

Notice that the lightest of the physical neutral scalars is a combination of $\left(\Phi\right)_{11}$, $\left(\Phi\right)_{22}$, $\chi_{kL}$ ($k=1,2$)
and should be interpreted as the SM-like 125 GeV Higgs particle found at the LHC. Furthermore, our model at low energies corresponds to a four Higgs doublet model with 2 scalar singlets coming from $\chi_{kR}$.  As seen from Eq. (\ref{LYU}), the top quark mass arises only from $\left( \Phi\right)_{11}$. Consequently, the dominant contribution to the SM-like 125 GeV Higgs arises mainly from $\left( \Phi\right)_{11}$. Here we note that there are enough free parameters in the low energy scalar potential to adjust the required pattern of scalar masses. This allows us to safely assume that the remaining scalars are heavy and outside the current reach of the LHC. In addition, the loop effects of the heavy scalars contributing to precision observables can be adequately suppressed by making an appropriate choice of the free parameters in the scalar potential. These adjustments do not affect the physical observables in the quark and lepton sectors, which are determined mainly by the Yukawa couplings.

We now explain the different group factors of the model. In the present model, the $\Delta \left( 27\right) $ group is responsible for the generation of a neutrino mass matrix texture compatible with the experimentally observed deviation of the tribimaximal mixing pattern. 
In addition it allows for renormalizable Yukawa terms only for the top quark, the gauge singlet Majorana fermions $\Omega_i$ ($i=1,2,3$) and tree level mass terms for the exotic charged fermions. This allows for their masses to appear at the tree level. Let us note that the $\Delta (27)$ discrete group is a non trivial group of the type $\Delta (3n^{2})$, isomorphic to the semi-direct product group $(Z_{3}^{\prime }\times Z_{3}^{\prime \prime})\times Z_{3}$ \cite{Ishimori:2010au}. This group was proposed for the first time in Ref. \cite{Branco:1983tn} and it has been employed in order to construct the Pati-Salam electroweak extension proposed in~\cite{CarcamoHernandez:2017owh}.
This group has also been used in multiscalar singlet models \cite{Bernal:2017xat}, multi-Higgs doublet models \cite{Bhattacharyya:2012pi,Aranda:2013gga}, Higgs triplet models \cite{CarcamoHernandez:2018djj}  SO(10) models~\cite{Bjorkeroth:2015uou,deMedeirosVarzielas:2017sdv,deMedeirosVarzielas:2018vab}, warped extra dimensional models \cite{Chen:2015jta}, and models based on the $SU(3)_{C}\otimes SU(3)_{L}\otimes U(1)_{X}$ gauge symmetry~\cite{Vien:2016tmh,Hernandez:2016eod,CarcamoHernandez:2018iel}. 

The auxiliary $Z_{6}$ and $Z_{12}$ symmetries select the allowed entries of the charged fermion mass matrices and shape their hierarchical structure, so as to get realistic SM charged lepton masses as well as quark mixing out of order one parameters. We assume that the $Z_{12}$ symmetry is broken down to a preserved $Z_{2}$ symmetry, which allows the implementation of an inverse/linear seesaw mechanism~\cite{Mohapatra:1986bd,GonzalezGarcia:1988rw,Akhmedov:1995vm,Akhmedov:1995ip,Malinsky:2005bi} for the generation of the light active neutrino masses. This is triggered by one-loop-induced seed mass parameters, in a manner analogous to the models discussed in~\cite{Bazzocchi:2009kc,CarcamoHernandez:2017kra}.
 
The spontaneously broken $Z_{4}$ symmetry also ensures the radiative nature of the inverse seessaw mechanism. This group was previously used in some other flavor
models and proved to be helpful, in particular, in the context of Grand Unification \cite{Emmanuel-Costa:2013gia,Arbelaez:2015toa,CarcamoHernandez:2018aon}, models with
extended $SU(3)_{C}\otimes SU(3)_{L}\otimes U(1)_{X}$ gauge symmetry \cite{Hernandez:2014vta,CarcamoHernandez:2017cwi}, extension of the inert doublet model \cite{CarcamoHernandez:2019cbd} and warped extra-dimensional models \cite{Hernandez:2015zeh}. It is worth mentioning that one or both of the $Z_{6}$ and $Z_{12}$ discrete groups were previously used in some other flavor models and proved to be useful in describing the SM fermion mass and mixing pattern, in particular in the context of two and three Higgs doublet models \cite{Arbelaez:2016mhg,Campos:2014zaa}, models with extended $SU(3)_{C}\otimes SU(3)_{L}\otimes U(1)_{X}$ gauge symmetry \cite{Hernandez:2014vta,Hernandez:2015tna,CarcamoHernandez:2017kra}, Grand Unified theories \cite{Arbelaez:2015toa} and models with strongly coupled heavy vector resonances \cite{CarcamoHernandez:2018vdj}.\\

Despite its extended particle spectrum, our model is minimal in the sense that each introduced field plays its own role in predicting viable fermion mass matrix textures that give rise to the observed SM fermion mass spectrum as well as fermion mixing parameters. This is achieved without the need to introduce 
hierarchy between the Yukawa couplings and heavy exotic fermion masses. The role of the different particles of our model is explained in the
following:
\begin{enumerate}
  \item The heavy exotic vector-like quarks $T_{k}$ ($k=1,2$), $B_{i}$ and heavy exotic charged leptons $E_{i}$ ($i=1,2,3$) represent the minimal set of exotic charged degrees of freedom needed to generate the masses for the up, charm, down, strange and bottom quarks and charged leptons via a Universal Seesaw mechanism. 
Note that for each SM charged fermion lighter than the top quark, one needs one exotic vector-like charged fermion to mediate the Universal Seesaw mechanism responsible for its mass. A reduced set of charged exotic fermions would lead to a proportionality between the rows and columns of the SM charged fermion mass matrix resulting
after the implementation of the Universal Seesaw mechanism, thus giving rise to a vanishing determinant for this matrix.
\item The scalar bi-doublet $\Phi $ is needed in order to build the renormalizable Yukawa term that gives rise to the tree-level top quark mass.
\item The $SU(2)_{L}$ doublet $\chi _{1L}$ and $SU(2)_{R}$ doublet $\chi _{1R}$ are crucial for generating the up, charm, down and strange quark masses, as well as the
masses of the charged leptons and the Cabbibo angle $\theta_{12}^{\left( q\right) }$.
\item The $SU(2)_{L}$ and $SU(2)_{R}$ doublets $\chi _{2L}$ and $\chi _{2R}$, respectively, are crucial to generate the bottom quark mass, and the 
quark mixing parameters $\sin\theta _{23}^{\left( q\right) }$ and $\sin \theta _{13}^{\left( q\right) }$
 as well as the quark CP violating phase $\delta $.
\item The gauge singlet scalar field $\sigma $ is required to generate a Froggatt-Nielsen picture of the CKM quark mixing matrix, crucial for
naturally explaining the observed hierarchies of the quark mixing parameters, in terms of powers of the Wolfenstein parameter $\lambda =0.225$. 
In addition, the singlet scalar field $\sigma $ is crucial for generating the Cabbibo-sized corrections to the leptonic mixing parameters (whose
main contribution arises from the light active neutrino mass matrix). This is needed to generate a realistic leptonic mixing pattern.
\item The scalar singlet $\eta $ is needed to account for the SM charged fermion mass hierarchy. Note that, despite the presence of several heavy vector-like 
fermions to trigger the Universal Seesaw mechanism, we assume that all of these have masses of the same order of magnitude, thus implying the need of implementing 
a Froggat-Nielsen mechanism. This, in combination with the Universal Seesaw mechanism, yields the SM charged fermion mass and quark mixing pattern. 
\item The $\Delta \left( 27\right) $ anti-triplet gauge singlet scalars $\rho $, $\phi $ and $\tau $ are crucial to make diagonal the heavy matrix blocks
associated to the Dirac neutrino states. This makes the light active neutrino mass matrix arising from the inverse seesaw, directly proportional
to the submatrix $\mu $, which characterizes the violation of lepton number by two units. On the other hand the $\Delta \left( 27\right) $ triplet
gauge singlet scalar $\xi $ is necessary to generate the Dirac neutrino mass submatrix, which is diagonal, due to the symmetries of our model. 
Thus, a realistic and predictive light active neutrino mass matrix exhibiting a generalized $\mu -\tau $ symmetry emerges, thanks to the presence of the 
$\Delta \left( 27\right) $ triplet gauge singlet scalar $\xi $. This also provides TeV-scale masses for the gauge singlet right handed Majorana
neutrinos $\Omega _{i}$ ($i=1,2,3$) that mediate, together with scalar singlet $\varphi $, the one-loop level inverse seesaw mechanism for the
generation of the light active neutrino masses. 
\item The gauge singlet scalar $\varphi $ is the only scalar in the model, which does not acquire vacuum expectation value. Its inclusion is crucial
for the implementation of the radiative inverse seesaw mechanism (at one-loop level) that produces small masses for the light active neutrinos.
\end{enumerate}

As seen from Eqs. (\ref{LYU}), (\ref{LYD}), (\ref{LYE}) and (\ref{LYnu}), we introduced several non-renormalizable Yukawa operators. 
These allow us to explain the observed hierarchies in the SM fermion mass spectrum and the fermion mixing parameters while keeping all the Yukawa 
couplings of order unity and the exotic fermion masses around the same order of magnitude. We now comment on the possible ultraviolet origin of these 
non-renormalizable operators.
%
%
Notice that all of them have the following form:
\begin{equation}
\overline{f}_{L}S_1F_R\left(\frac{\Sigma_1}{\Lambda}\right)^{n_1}\hspace{1cm}\overline{F}_{L}S_2f_R\left(\frac{\Sigma_2}{\Lambda}\right)^{n_1}
\label{NRoperators}
\end{equation}
where $f$ and $F$ stand for light and heavy fermions, respectively, $n_1$, $n_2$ are integers and $S_1$, $S_2$, $\Sigma_1$ and $\Sigma_2$ are scalars. Here, 
for simplicity, we have omitted family and fermionic type indices. One sees that these non-renormalizable operators in Eq. (\ref{NRoperators}) can 
all arise from the following renormalizable operators:
\begin{eqnarray}
&&\overline{f}_{L}S_3\tilde{F}_R\hspace{1cm}\overline{\tilde{F}}_{L}S_4F_R\hspace{1cm}\overline{\tilde{F}}_{L}S_5\tilde{F}_R\notag\\
&&\overline{f}_{L}S_6\tilde{F}_R\hspace{1cm}\overline{\tilde{F}}_{L}S_7F_R\hspace{1cm}\overline{\tilde{F}}_{L}S_8\tilde{F}_R
\label{Roperators}
\end{eqnarray}
where $S_k$ ($k=3,4,\cdots 8$) are extra scalars and $\tilde{F}$ extra very heavy fermions. Assuming that the $S_5$ and $S_8$ scalars acquire vacuum 
expectation values much larger than the remaining scalars, the fermions $\tilde{F}$ will get very large masses. As a result, they can be integrated out, 
yielding effective non-renormalizable operators as in Eq.~(\ref{NRoperators}).

Quark masses and mixing parameters are modeled with the help of the scalar singlets $\sigma $ and $\eta $. We assume that these scalars acquire vacuum expectation values 
of order $\lambda \Lambda $, where $\lambda =0.225$ is the Cabibbo angle and $\Lambda $ is the cutoff of our model. Consequently, we set the VEVs of the scalar fields to satisfy the following hierarchy:
\begin{equation}
v_{1} \sim v_{kL}\sim v\ll v_{kR}\sim v_{\xi }\ll v_{\rho }\sim v_{\phi }\sim v_{\tau
}\sim v_{\eta }\sim v_{\sigma }\sim\lambda \Lambda ,\hspace{1.5cm}k=1,2.
\label{VEVhierarchy}
\end{equation}%
Here $v=246$ GeV is the electroweak symmetry breaking scale and $v_{kR}\gtrsim$ $\mathcal{O}(10)$ TeV ($k=1,2$) the scale of breaking of the left-right symmetry. 
It has been shown in detail in Ref. \cite{Chauhan:2018uuy} that this lower bound on the scale of breaking of the $SU(2)_R$ symmetry is obtained from stringent flavor constraints, assuming that all the gauge and quartic couplings remain perturbative up to the GUT scale.
This makes it hard to detect the  $SU(2)_R$ gauge bosons at the LHC. However, a future $100$ TeV proton-proton collider could probe signatures associated to $W_R$ and $Z_R$ gauge bosons and hence test the model. Moreover, our model can also be tested via the production and decays of the heavy vector-like fermions that mediate the Universal Seesaw Mechanism. As seen from Eqs. (\ref{LYU}), (\ref{LYD}) and (\ref{LYE}), the masses of these exotic fermions are not related to the $SU(2)_R$ symmetry breaking scale could be of the order of few TeV, making them potentially accesible to LHC searches. 
A detailed study of the collider phenomenology of our model goes beyond the scope of this work and is deferred for future studies.

The resulting mixing angles of $\xi $ with $\rho $, $\eta $ and $\tau $ are very tiny since they are suppressed by the ratios of their VEVs, which is a consequence of the method of recursive expansion proposed in Ref. \cite{Grimus:2000vj}. Thus, the scalar potential for $\xi $ can be studied independently from the corresponding one for $\rho$, $\eta $, $\tau $. As shown in detail in Ref. \cite{CarcamoHernandez:2017owh}, the following VEV alignments for the $\Delta (27) $ scalar triplets are consistent with the scalar potential minimization equations for a large region of parameter space: 
\begin{equation}
\vev{ \rho} =v_{\rho }\left( 1,0,0\right) ,\hspace{1.5cm}%
\vev{ \phi} =v_{\phi }\left( 0,1,0\right) ,\hspace{1.5cm}%
\vev{ \tau} =v_{\tau }\left( 0,0,1\right) ,\hspace{1.5cm}%
\vev{ \xi} =\frac{v_{\xi }}{\sqrt{2+r^{2}}}\left(
r,e^{-i\psi },e^{i\psi }\right) .
\end{equation}

Summarizing, the full symmetry of the model exhibits the following
spontaneous breaking pattern
\begin{gather}
\mathcal{G}=SU(3)_{C}\otimes SU\left( 2\right) _{L}\otimes SU\left(
2\right) _{R}\otimes U\left( 1\right) _{B-L}\otimes \Delta \left( 27\right)
\otimes Z_{4}\otimes Z_{6}\otimes Z_{12}  \label{Group} \\
\Downarrow \Lambda _{int}  \notag \\SU(3)_{C}\otimes SU\left( 2\right) _{L}\otimes SU\left(
2\right) _{R}\otimes U\left( 1\right) _{B-L}\otimes Z_{4}\otimes Z_{2} \notag \\\Downarrow v_{kR},v_{\xi }  \notag \\SU(3)_{C}\otimes SU\left( 2\right) _{L}\otimes U\left(
1\right) _{Y}\otimes Z_{2}  \notag \\\Downarrow v_{1},v_{kL}  \notag \\SU(3)_{C}\otimes U\left( 1\right) _{Q}\otimes Z_2.  \notag
\end{gather}  

\section{Lepton masses and mixings}

\label{leptonmassesandmixing}
\subsection{Charged lepton sector}

From the charged lepton Yukawa terms in Eq.~(\ref{LYE}), we find that the mass matrix containing the charged leptons in the basis $(\overline{l}_{1L},\overline{l}_{2L},\overline{l}_{3L},\overline{E}_{1L},\overline{E}_{2L},\overline{E}_{3L})$ versus $(l_{1R},l_{2R},l_{3R},E_{1R},E_{2R},E_{3R})$ takes the form: 
\begin{equation*}
M_{E}=\left( 
\begin{array}{cc}
0_{3\times 3} & z\frac{v_{L}}{\sqrt{2}} \\ 
z^{T}\frac{v_{R}}{\sqrt{2}} & m_{E}%
\end{array}%
\right) ,\hspace{1cm}z=\left( 
\begin{array}{ccc}
z_{11}\lambda ^{2} & 0 & z_{13}\lambda ^{2} \\ 
0 & z_{22}\lambda &z_{23}\lambda\\ 
0 & 0 & z_{33}%
\end{array}%
\right) \allowbreak ,\hspace{1cm}m_{E}=\left( 
\begin{array}{ccc}
m_{E_{1}} & 0 & 0 \\ 
0 & m_{E_{2}} & 0 \\ 
0 & 0 & m_{E_{3}}%
\end{array}%
\right) \allowbreak \allowbreak
\end{equation*}

Given that the exotic charged lepton masses $m_{E_{i}}$ ($i=1,2,3$) are much larger than $v_{L}$ and $v_{R}$,  it follows that the SM charged leptons get their masses from an Universal seesaw mechanism mediated by the three charged exotic leptons $E_{i}$ ($i=1,2,3$). Then, the SM charged lepton mass matrix becomes 
\begin{eqnarray}\label{chlm}
\widetilde{M}_{E} &=&\frac{v_{L}v_{R}}{2}zm_{E}^{-1}z^{T}=\left( 
\begin{array}{ccc}
\left( \frac{m_{E_{3}}}{m_{E_{1}}}z_{11}^{2}+z_{13}^{2}\right) \lambda ^{4}
& z_{13}z_{23}\lambda ^{3} & z_{13}z_{33}\lambda ^{2} \\ 
z_{13}z_{23}\lambda ^{3} & \left( \frac{m_{E_{3}}}{m_{E_{2}}}z_{22}^{2}+z_{23}^{2}\right) \lambda ^{2} & z_{23}z_{33}\lambda  \\ 
z_{13}z_{33}\lambda ^{2} & z_{23}z_{33}\lambda  & z_{33}^{2}\end{array}\right) \allowbreak \frac{v_{L}v_{R}}{2m_{E_{3}}}  \notag \\
&=&\left( 
\begin{array}{ccc}
e_{11}\lambda ^{7} & e_{12}\lambda ^{6} & e_{13}\lambda ^{5} \\ 
e_{12}\lambda ^{6} & e_{22}\lambda ^{5} & e_{23}\lambda ^{4} \\ 
e_{13}\lambda ^{5} & e_{23}\lambda ^{4} & e_{33}\lambda ^{3}\end{array}\right) \allowbreak \frac{v}{\sqrt{2}},
\end{eqnarray}
where the effective Yukawas $e_{ij}$ are naturally expected to be of order one. The \sm charged lepton mass matrix is diagonalized by a unitary matrix through 
$V_{l}^{\dagger } \widetilde{M}_{E}\widetilde{M}_{E}^{\dagger }V_{l}=\mathrm{diag} (m_{e}^{2},m_{\mu }^{2},m_{\tau }^{2})$. In order to illustrate how the charged lepton mass spectrum arises from Eq.(\ref{chlm}), we can choose the benchmark point 
\begin{align}
e_{11}& =e_{12}=e_{22}=e_{23}=1.3622, & e_{13}& =1.61464, & e_{33}&
=0.759133,\,
\end{align}%
assuming real entries in $\widetilde{M}_{E}$, to get the mass eigenvalues $m_{e}=0.487\,\mathrm{MeV}$, $m_{\mu }=102.8\,\mathrm{MeV}$ , $m_{\tau }=1.75\,\mathrm{GeV}$ with  the charged lepton mixing matrix
\begin{equation}\label{chm}
V_{l}=\left( 
\begin{array}{ccc}
0.952869 & -0.288859 & 0.0927446 \\ 
-0.302746 & -0.885564 & 0.352309 \\ 
0.0196361 & 0.363782 & 0.931277 
\end{array}
\right).
\end{equation}

\subsection{Neutrino sector}

\begin{figure}[tbh]
\begin{center}
\includegraphics[scale=0.8]{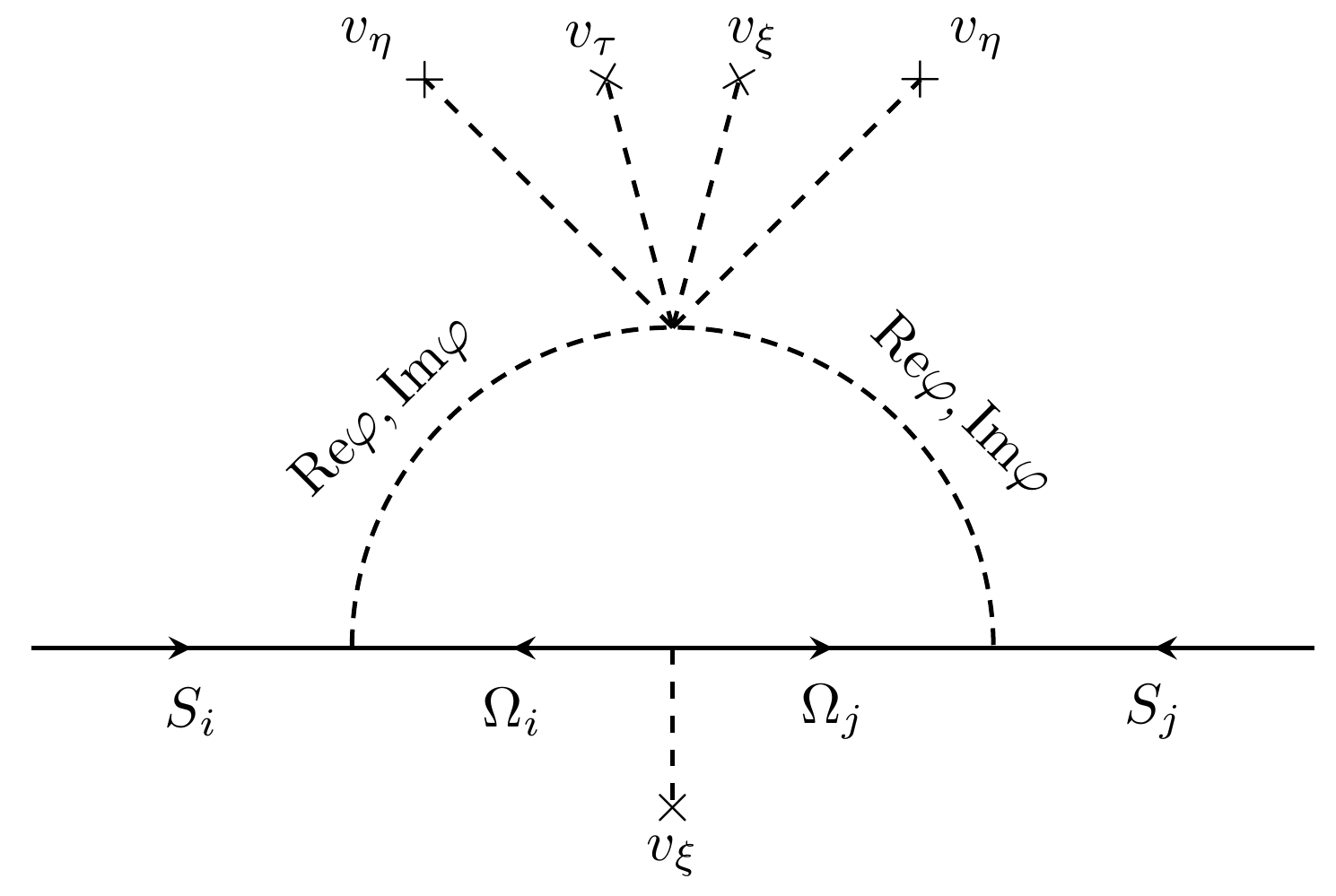}
\end{center}
\caption{Loop Feynman diagram contributing to the Majorana neutrino mass submatrix entries  $\protect\mu_{ij}$, with $i,j=1,2,3$.}
\label{Loopmu}
\end{figure}

From the neutrino Yukawa interactions in Eq.~(\ref{LYnu}), we obtain the following mass terms: 
\begin{equation}
-\mathcal{L}_{mass}^{\left( \nu \right) }=\frac{1}{2}\left( 
\begin{array}{ccc}
\overline{\nu _{L}^{C}} & \overline{\nu _{R}} & \overline{S}%
\end{array}%
\right) M_{\nu }\left( 
\begin{array}{c}
\nu _{L} \\ 
\nu _{R}^{C} \\ 
S^{C}%
\end{array}%
\right) +H.c,  \label{Lnu}
\end{equation}%
where the neutrino mass matrix $M_{\nu }$ is given in block form as: 
\begin{eqnarray}
M_{\nu } &=&\left( 
\begin{array}{ccc}
0_{3\times 3} & M_{1} & \frac{v_{L}v_{\xi }^{2}}{v_{R}\Lambda ^{2}}M_{2} \\ 
M_{1}^{T} & 0_{3\times 3} & M_{2} \\ 
\frac{v_{L}v_{\xi }^{2}}{v_{R}\Lambda ^{2}}M_{2}^{T} & M_{2}^{T} & \mu%
\end{array}%
\right) ,\hspace{1cm}M_{1}=\left( 
\begin{array}{ccc}
\gamma _{1} & 0 & 0 \\ 
0 & \gamma _{2} & 0 \\ 
0 & 0 & \gamma _{3}%
\end{array}%
\right) \allowbreak \frac{vv_{\xi }^{2}}{\sqrt{2}\Lambda ^{2}},\hspace{1cm}%
M_{2}=\left( 
\begin{array}{ccc}
\kappa _{1}\frac{v_{\rho }}{\Lambda } & 0 & 0 \\ 
0 & \kappa _{2}\frac{v_{\phi }}{\Lambda } & 0 \\ 
0 & 0 & \kappa _{3}\frac{v_{\tau }}{\Lambda }%
\end{array}%
\right) \allowbreak \frac{v_{R}v_{\sigma }}{\sqrt{2}\Lambda },  \notag \\
\mu &=&\left( 
\begin{array}{ccc}
r\lambda _{1}F\left( \frac{\lambda _{1}v_{\xi }}{\sqrt{2+r^{2}}}%
,m_{R},m_{I}\right) & \lambda _{2}e^{i\psi }F\left( \frac{\lambda
_{2}e^{i\psi }v_{\xi }}{\sqrt{2+r^{2}}},m_{R},m_{I}\right) & \lambda
_{2}e^{-i\psi }F\left( \frac{\lambda _{2}e^{-i\psi }v_{\xi }}{\sqrt{2+r^{2}}}%
,m_{R},m_{I}\right) \\ 
&  &  \\ 
\lambda _{2}e^{i\psi }F\left( \frac{\lambda _{2}e^{i\psi }v_{\xi }}{\sqrt{%
2+r^{2}}},m_{R},m_{I}\right) & \lambda _{1}e^{-i\psi }F\left( \frac{\lambda
_{1}e^{-i\psi }v_{\xi }}{\sqrt{2+r^{2}}},m_{R},m_{I}\right) & r\lambda
_{2}F\left( \frac{r\lambda _{2}v_{\xi }}{\sqrt{2+r^{2}}},m_{R},m_{I}\right)
\\ 
&  &  \\ 
\lambda _{2}e^{-i\psi }F\left( \frac{\lambda _{2}e^{-i\psi }v_{\xi }}{\sqrt{%
2+r^{2}}},m_{R},m_{I}\right) & r\lambda _{2}F\left( \frac{r\lambda
_{2}v_{\xi }}{\sqrt{2+r^{2}}},m_{R},m_{I}\right) & \lambda _{1}e^{i\psi
}F\left( \frac{\lambda _{1}e^{i\psi }v_{\xi }}{\sqrt{2+r^{2}}}%
,m_{R},m_{I}\right)%
\end{array}%
\right) \allowbreak \frac{v_{\xi }}{\sqrt{2+r^{2}}},
\end{eqnarray}%
with $m_{R}=m_{\func{Re}\varphi }$ and $m_{I}=m_{\func{Im}\varphi }$ and the loop function \cite{Ma:2006km}: 
\begin{equation}
F\left( m_{1},m_{2},m_{3}\right) =\frac{\lambda _{3}^{2}}{16\pi ^{2}}\left[ 
\frac{m_{2}^{2}}{m_{2}^{2}-m_{1}^{2}}\ln \left( \frac{m_{2}^{2}}{m_{1}^{2}}%
\right) -\frac{m_{3}^{2}}{m_{3}^{2}-m_{1}^{2}}\ln \left( \frac{m_{3}^{2}}{%
m_{1}^{2}}\right) \right] .
\end{equation}%
The one-loop Feynman diagrams contributing to the entries of the Majorana neutrino mass submatrix $\mu $ are shown in Fig.~\ref{Loopmu}. The splitting between the masses $m_{\func{Re}\varphi }$ and $m_{\func{Im}\varphi }$ arises from the term 
\begin{equation}
\frac{\kappa }{\Lambda ^{2}}\left( \varphi ^{\ast }\right) ^{2}\tau ^{\ast
}\xi ^{\ast }\left( \eta ^{\ast }\right) ^{2}.
\end{equation}%
For the sake of simplicity, we assume that the singlet scalar field $\varphi $\ is heavier than the right-handed Majorana neutrinos $\Omega _{i}$ ($i=1,2,3$), so that we can restrict to the scenario 
\begin{equation}
m_{R}^{2},m_{I}^{2} \gg \max \left( \frac{\lambda _{1}^{2}\allowbreak v_{\xi}^{2}}{2+r^{2}},\frac{r^{2}\lambda _{1}^{2}\allowbreak v_{\xi }^{2}}{2+r^{2}}%
\right) ,
\end{equation}%
and $m_{R}^{2}-m_{I}^{2} \ll m_{R}^{2}+m_{I}^{2}$, for which the submatrix $\mu$ takes the form 
\begin{equation}
\mu \simeq \frac{\lambda _{3}^{2}\left( m_{R}^{2}-m_{I}^{2}\right) v_{\xi }}{%
8\pi ^{2}\left( m_{R}^{2}+m_{I}^{2}\right) \sqrt{2+r^{2}}}\left( 
\begin{array}{ccc}
r\lambda _{1} & \lambda _{2}e^{i\psi } & \lambda _{2}e^{-i\psi } \\ 
&  &  \\ 
\lambda _{2}e^{i\psi } & \lambda _{1}e^{-i\psi } & r\lambda _{2} \\ 
&  &  \\ 
\lambda _{2}e^{-i\psi } & r\lambda _{2} & \lambda _{1}e^{i\psi }%
\end{array}%
\right) \allowbreak .
\end{equation}

The structure of the resulting neutrino mass is a particular case of that in Ref.~\cite{Grimus:2003yn} (see below).
Besides the three active Majorana neutrinos the physical states include the six heavy exotic neutrinos.
After seesaw block-diagonalization~\cite{Schechter:1981cv} we obtain
\begin{equation}
M_{\nu }^{\left( 1\right) }=\left( \frac{vv_{\xi }^{2}}{v_{\rho
}v_{\sigma }v_{R}}\right) ^{2}\mu -\frac{v_{L}v_{\xi }^{2}}{v_{R}\Lambda
^{2}}\left( M_{1}+M_{1}^{T}\right),\hspace{.5cm}M_{\nu
}^{\left( 2\right) }=-\frac{1}{2}\left( M_{2}+M_{2}^{T}\right)
+\frac{1}{2}\mu ,\hspace{.5cm}M_{\nu }^{\left( 3\right) }=\frac{1}{2}\left(
M_{2}+M_{2}^{T}\right) +\frac{1}{2}\mu .  \label{Mnu1}
\end{equation}
where we have simplified our analysis setting $\gamma_{i}=\kappa _{i}$ ($i=1,2,3$). Here $M_{\nu }^{\left(1\right) }$ corresponds to the effective active neutrino mass matrix resulting from seesaw diagonalization, whereas $M_{\nu}^{\left( 2\right) }$ and $M_{\nu }^{\left( 3\right) }$ correspond to the heavy blocks associated to the exotic Dirac states. These form three quasi-Dirac pairs that can lie at the TeV scale, with a small splitting $\mu $. The first term in $M_{\nu }^{\left( 1\right) }$ corresponds to the inverse seesaw piece~\cite{Mohapatra:1986bd,GonzalezGarcia:1988rw}, while the latter comes from the linear seesaw contribution~\cite{Akhmedov:1995ip,Akhmedov:1995vm,Malinsky:2005bi}.

The light neutrino mass matrix can be further simplified ignoring the diagonal contributions from the linear seesaw term ($v_L \ll v_R$). In this approximation, the mass matrix becomes simply $M_{\nu }^{\left( 1\right) }=\mu$. Taking real Yukawa couplings $\lambda_i$ and VEVs, we find that $M_{\nu }^{\left( 1\right) }$ has explicit generalized $\mu-\tau$ symmetry~\cite{babu:2002dz,Grimus:2003yn,King:2014nza}
\begin{equation}
X^{T}M_{\nu }^{\left( 1\right) } X =M_{\nu }^{\left( 1\right)* },
\end{equation}
with 
\begin{equation}
X=\left( 
\begin{array}{ccc}
1 & 0 & 0 \\ 
0 & 0 & 1 \\ 
0 & 1 & 0%
\end{array}
\right).
\end{equation}
The most general matrix $V_{\nu}$ that diagonalizes $M_{\nu }^{\left(
1\right) }$ with $V_{\nu}^{T}M_{\nu }^{\left( 1\right) }V_{\nu}=\mathrm{diag}%
(m^{\nu}_{1},m^{\nu}_{2},m^{\nu}_{3})$ can be written as \cite%
{Chen:2015siy,Chen:2016ica} 
\begin{equation}
V_{\nu}=\Sigma O_{23} O_{13} O_{12}Q_{\nu},
\end{equation}
where the matrix
\begin{equation}
\Sigma=\left( 
\begin{array}{ccc}
1 & 0 & 0 \\ 
0 & \frac{1}{\sqrt{2}} & \frac{i}{\sqrt{2}} \\ 
0 & \frac{1}{\sqrt{2}} & -\frac{i}{\sqrt{2}}%
\end{array}
\right),
\end{equation}
stands for the Takagi factorization of $X$, satisfying $X=\Sigma \Sigma^{T}$. The $O_{ij}$ are $3\times 3$ orthogonal matrices parameterized as 
\begin{equation}  \label{Ort}
\begin{split}
O_{23}&=\left( 
\begin{array}{ccc}
1 & 0 & 0 \\ 
0 & \cos \omega_{23} & \sin \omega_{23} \\ 
0 & -\sin \omega_{23} & \cos \omega_{23}%
\end{array}
\right)\,,\qquad O_{13}= \left( 
\begin{array}{ccc}
\cos \omega_{13} & 0 & \sin \omega_{13} \\ 
0 & 1 & 0 \\ 
-\sin \omega_{13} & 0 & \cos \omega_{13}%
\end{array}
\right)\,, \\
\qquad O_{12}&= \left( 
\begin{array}{ccc}
\cos \omega_{12} & \sin \omega_{12} & 0 \\ 
-\sin \omega_{12} & \cos \omega_{12} & 0 \\ 
0 & 0 & 1%
\end{array}
\right),
\end{split}%
\end{equation}
and $Q_{\nu}=\mathrm{diag}(e^{- i \pi k_1/2},e^{- i \pi k_2/2},e^{- i \pi k_3/2})$ is a diagonal matrix of phases, with $k_i=0,1,2,3$.

Due to the reduced number of parameters in our model, we have the following relations among the parameters of the mixing matrix: 
\begin{equation}  \label{pr}
\begin{split}
&\tan2\omega_{12}=\frac{4 \sin \omega_{13}\left[\sin \left(2 \psi +\omega
_{23}\right)+\sin 3 \omega_{23}\right]}{\left(3 \cos 2 \omega _{13}-1\right)
\cos \left(2 \psi +\omega_{23}\right)+\left(\cos 2 \omega _{13}-3\right)
\cos 3 \omega_{23}}\,, \\
&\lambda_1=-\sqrt{2} \lambda_2 \tan \omega_{13} \cos \left(\psi -\omega
_{23}\right) \csc \left(\psi +2 \omega_{23}\right) \\
&r=-\frac{\sqrt{2} \left[\sin \omega_{13} \cos \left(\psi -\omega
_{23}\right) \cos \left(\psi +2 \omega_{23}\right)+2 \cos \omega_{13} \cot 2
\omega_{13} \sin \left(\psi -\omega_{23}\right) \sin \left(\psi +2
\omega_{23}\right)\right]}{\sqrt{2} \sin \omega_{13} \cos \left(\psi
-\omega_{23}\right)+\cos \omega_{13} \sin \left(\psi +2 \omega_{23}\right)}%
\,.
\end{split}%
\end{equation}
Finally, the light active neutrino mass spectrum is 
\begin{equation}
\mathrm{diag}
(m^{\nu}_{1},m^{\nu}_{2},m^{\nu}_{3})=Q_{\nu}^{T}\mathrm{diag}(m_{1},m_{2},m_{3})Q_{\nu},
\end{equation}
with positive definite physical masses $m^{\nu}_{i}$, $i=1,2,3$, and
\begin{equation}
\begin{split}
m_{1}=& m_0\bigg\{\frac{\sqrt{2}%
\tan \omega_{13}\sin 2\left( \psi -\omega_{23}\right) +\left( 1-4\tan
^{2}\omega_{13}\right) \cos \left( 2\psi +\omega_{23}\right) -\cos 3\omega
_{23}}{2\tan \omega_{13}\left[ \sqrt{2}\sin \left( \psi +2\omega
_{23}\right) +2\tan \omega_{13}\cos \left( \psi -\omega_{23}\right) \right]
} \\
& \qquad -\frac{\left( 3\cos 2\omega_{13}-1\right) \cos \left( 2\psi
+\omega_{23}\right) +\left( \cos 2\omega_{13}-3\right) \cos 3\omega_{23}}{%
2\sqrt{2}\sin 2\omega_{13}\sin \left( \psi +2\omega_{23}\right) } \\
& \qquad \times \sqrt{1+\frac{16\sin ^{2}\omega_{13}\left[ \sin \left(
2\psi +\omega_{23}\right) +\sin \left( 3\omega_{23}\right) \right] {}^{2}}{%
\left[ \left( 3\cos 2\omega_{13}-1\right) \cos \left( 2\psi +\omega
_{23}\right) +\left( \cos 2\omega_{13}-3\right) \cos 3\omega_{23}\right]
{}^{2}}}\bigg\}\,, \\
m_{2}=& m_0\bigg\{\frac{\sqrt{2}%
\tan \omega_{13}\sin 2\left( \psi -\omega_{23}\right) +\left( 1-4\tan
^{2}\omega_{13}\right) \cos \left( 2\psi +\omega_{23}\right) -\cos 3\omega
_{23}}{2\tan \omega_{13}\left[ \sqrt{2}\sin \left( \psi +2\omega
_{23}\right) +2\tan \omega_{13}\cos \left( \psi -\omega_{23}\right) \right]
} \\
& \qquad +\frac{\left( 3\cos 2\omega_{13}-1\right) \cos \left( 2\psi
+\omega_{23}\right) +\left( \cos 2\omega_{13}-3\right) \cos 3\omega_{23}}{%
2\sqrt{2}\sin 2\omega_{13}\sin \left( \psi +2\omega_{23}\right) } \\
& \qquad \times \sqrt{1+\frac{16\sin ^{2}\omega_{13}\left[ \sin \left(
2\psi +\omega_{23}\right) +\sin \left( 3\omega_{23}\right) \right] {}^{2}}{%
\left[ \left( 3\cos 2\omega_{13}-1\right) \cos \left( 2\psi +\omega
_{23}\right) +\left( \cos 2\omega_{13}-3\right) \cos 3\omega_{23}\right]
{}^{2}}}\bigg\}\,, \\
m_{3}=& m_0\bigg\{\frac{\tan
^{2}\omega_{13}\left[ \cos \left( \psi +2\omega_{23}\right) +\cos 3\psi %
\right] \csc ^{2}\left( \psi +2\omega_{23}\right) -\sqrt{2}\cot \omega
_{13}\sin \left( \psi -\omega_{23}\right) }{1+\sqrt{2}\tan \omega_{13}\cos
\left( \psi -\omega_{23}\right) \csc \left( \psi +2\omega_{23}\right) }%
\bigg\}\,,
\end{split}
\end{equation}
in terms of a common mass scale $m_0$
\begin{equation}
m_0=   \frac{\lambda_2\lambda _{3}^{2}\left( m_{R}^{2}-m_{I}^{2}\right) v_{\xi }}{%
8\pi ^{2}\sqrt{2+r^{2}}\left( m_{R}^{2}+m_{I}^{2}\right) }\left( \frac{vv_{\xi }^{2}}{v_{\rho
}v_{\sigma }v_{R}}\right) ^{2}\approx\frac{\lambda_2\lambda _{3}^{2}\left( m_{R}^{2}-m_{I}^{2}\right)v_{\xi }}{%
8\pi ^{2}\lambda^4\sqrt{2+r^{2}}\left( m_{R}^{2}+m_{I}^{2}\right) }\left( \frac{v^2v_{\xi }^{2}}{ \Lambda^4}\right).
\end{equation}

\subsection{Lepton mixing matrix}

The lepton mixing matrix is thus given by 
\begin{equation}  \label{LMM}
U=V_{l}^{\dagger}V_{\nu},
\end{equation}
where we take $V_{l}$ to be approximated as
\begin{equation}
V_{l}\approx\left(
\begin{array}{ccc}
 \cos \eta_1 & \sin \eta_1 & 0 \\
 -\sin \eta_1 & \cos \eta_1 & 0 \\
 0 & 0 & 1 \\
\end{array}
\right)\left(
\begin{array}{ccc}
 1 & 0 & 0 \\
 0 & \cos \eta_2 & \sin \eta_2 \\
 0 & -\sin \eta_2 & \cos \eta_2 \\
\end{array}
\right)=\left(
\begin{array}{ccc}
 \cos \eta _1 & \cos\eta _2 \sin\eta _1 & \sin
   \eta _1 \sin\eta _2 \\
 -\sin \eta _1& \cos\eta _1 \cos \eta _2 & \cos
   \eta _1 \sin\eta _2\\
 0 & -\sin \eta _2 & \cos \eta _2
\end{array}
\right),
\end{equation}
with $\eta_1$ and $\eta_2$ of the same order as the Cabibbo parameter $\lambda$, as indicated by our estimate in Eq.(\ref{chm}).
In the fully ``symmetrical'' presentation of the lepton mixing matrix~\cite{Schechter:1980gr,Rodejohann:2011vc}  
\begin{equation}  \label{eq:symmetric_para}
U = \left( 
\begin{array}{ccc}
c_{12} c_{13} & s_{12} c_{13} e^{ - i \phi_{12} } & s_{13} e^{ -i \phi_{13} }
\\ 
-s_{12} c_{23} e^{ i \phi_{12} } - c_{12} s_{13} s_{23} e^{ -i ( \phi_{23} -
\phi_{13} ) } & c_{12} c_{23} - s_{12} s_{13} s_{23} e^{ -i ( \phi_{23} +
\phi_{12} - \phi_{13} ) } & c_{13} s_{23} e^{- i \phi_{23} } \\ 
s_{12} s_{23} e^{ i ( \phi_{23} + \phi_{12} ) } - c_{12} s_{13} c_{23} e^{ i
\phi_{13} } & - c_{12} s_{23} e^{ i \phi_{23} } - s_{12} s_{13} c_{23} e^{
-i ( \phi_{12} - \phi_{13} ) } & c_{13} c_{23}%
\end{array}
\right)\,,
\end{equation}
with $c_{ij}=\cos\theta_{ij}$ and $s_{ij}=\sin\theta_{ij}$, we find the relation between mixing angles and the entries of $U$ to be given as 
\begin{equation}  \label{eq:UU}
\sin^{2} \theta_{13} = \left| U_{e3} \right|^{2} \, , \quad \sin^{2}
\theta_{12} = \frac{ \left| U_{e2} \right|^{2} }{ 1 - \left| U_{e3}
\right|^{2} } \quad \text{and} \quad \sin^{2} \theta_{23} = \frac{ \left|
U_{\mu 3} \right|^{2} }{ 1 - \left| U_{e3} \right|^{2} } \,.
\end{equation}
The Jarlskog invariant $J_{\mathrm{CP}}=\mathrm{Im}\left(
U_{11}^{*} U_{23}^{*} U_{13} U_{21} \right)$, 
takes the form~\cite{Rodejohann:2011vc} 
\begin{equation}
J_{\mathrm{CP}} = \frac{1}{8} \sin 2 \theta_{12} \, \sin 2 \theta_{23} \,
\sin 2 \theta_{13}\, \cos\theta_{13} \,\sin(\phi_{13}-\phi_{23}-\phi_{12})
\,,
\end{equation}
This is the CP phase relevant for the description of neutrino oscillations. The two additional Majorana-type rephasing invariants $I_{1} = \mathrm{Im}\left( U_{12}^{2} U_{11}^{* 2} \right)$ and $I_{2} = \mathrm{Im} \left( U_{13}^{2} U_{11}^{*2} \right)$ are given as
\begin{equation}
\begin{array}{l}
I_{1} = \frac{1}{4} \sin^{2} 2\theta_{12} \cos^{4} \theta_{13} \sin ( - 2
\phi_{12} ) \quad \text{and} \quad I_{2} = \frac{1}{4} \sin^{2} 2
\theta_{13} \cos^{2} \theta_{12} \sin ( - 2 \phi_{13} )\,.%
\end{array}%
\end{equation}

In terms of the model parameters, the lepton mixing angles are expressed as 
\begin{equation}  \label{s13}
\sin^2 \theta_{13}=\frac{1}{4} \left[2 \sin ^2\eta_1+(1+3 \cos 2 \eta_1) \sin
   ^2\omega _{13}-\sqrt{2} \sin2 \eta_1 \sin 2 \omega _{13} \sin \omega _{23}\right]\,,
\end{equation}
\begin{equation}  \label{s12}
\begin{split}
\sin^2 \theta_{12}=&\big[2 \sin ^2\eta_1 +(1+3 \cos 2 \eta_1 ) \sin ^2\omega _{12} \cos ^2 \omega _{13}\\&-2 \sqrt{2} \sin 2 \eta_1  \sin\omega _{12}
   \cos \omega _{13} \left(\cos \omega _{12} \cos \omega _{23}-\sin \omega _{12} \sin
   \omega _{13} \sin\omega _{23}\right)\big] \\
& \times \big[4-2 \sin ^2\eta_1-(1+3 \cos 2 \eta_1) \sin
   ^2\omega _{13}+\sqrt{2} \sin2 \eta_1 \sin 2 \omega _{13} \sin \omega _{23}\big]^{-1}\,,
\end{split}%
\end{equation}
\begin{equation}  \label{s23}
\begin{split}
&\sin^2 \theta_{23}=\bigg\{5+\cos 2 \omega _{13} \left(1+3 \cos 2 \eta _1-6 \sin ^2\eta
   _1 \cos 2 \eta _2\right)-\cos 2 \eta _1+2 \sin ^2\eta _1 \cos
   2 \eta _2\\&
\qquad   +8 \cos \eta _2\cos
  \omega _{13} \left[2 \sin \eta _2 \cos\eta _1 \cos
  \omega _{13} \cos 2 \omega _{23}+\sqrt{2} \sin\omega
   _{13} \sin\omega _{23} \left(\sin 2 \eta _1 \cos
 \eta _2-2 \sin \eta _1 \sin \eta
   _2\right)\right]\bigg\} \\
&\qquad   \times \big\{4\big[4-2 \sin ^2\eta_1-(1+3 \cos 2 \eta_1) \sin
   ^2\omega _{13}+\sqrt{2} \sin2 \eta_1 \sin 2 \omega _{13} \sin \omega _{23}\big]\big\}^{-1}\,,
\end{split}%
\end{equation}
leading to the correlations
\begin{equation}  \label{corr1}
\begin{split}
 \cos ^2\theta_{13}(\cos 2 \theta_{12}-\cos 2 \omega_{12} )=&-\cos 2 \omega_{12}\sin ^2\eta_1+\frac{1}{\sqrt{2}}\sin 2 \eta_1
\sin 2 \omega_{12} \cos \omega_{13} \cos \omega_{23},
\end{split}
\end{equation}
\begin{equation}  \label{corr2}
\begin{split}
\cos ^2\theta_{13} \cos 2\theta_{23}=&\frac{1}{4} \big\{\sin ^2\eta _1 \cos 2 \eta _2 \left(3
   \cos 2 \omega _{13}-1\right)-4 \sin2 \eta _2 \cos \eta _1 \cos 2 \omega _{23} \cos
   ^2\omega _{13}\\&+2 \sqrt{2} \sin \eta _1\sin 2
   \omega _{13} \sin\omega _{23} \left(\sin2 \eta _2-\cos
   \eta _1 \cos 2 \eta _2\right)\big\}.
   \end{split}
\end{equation}
The right-hand side of both relations vanishes in the limit  $\eta _1 ,\eta _2 \to 0$, recovering the predictions of generalized $\mu-\tau$ symmetry, namely, $ \theta_{12}= \omega_{12}$ and $\theta_{23}=\pi/4$. 

Analogously, the rephasing invariants can be written in terms of the model parameters as
\begin{equation}
\begin{split}
&J_{\mathrm{CP}}=\frac{1}{64} \Bigg\{\sin \eta
   _1 \bigg\{\sqrt{2} \big\{\sin 2 \eta _2 \cos ^2\eta _1
   \left[4 \sin2 \omega _{13} \cos2 \omega _{12} \cos3
   \omega _{23}+\sin2 \omega _{12} \sin3 \omega _{23}
   \left(\cos 3 \omega _{13}-5 \cos \omega
   _{13}\right)\right]\\
   &\quad-\left(2 \cos \eta _1 \cos 2 \eta
   _2+\sin 2 \eta _2 \cos 2 \eta _1\right) \left[4 \sin
   2 \omega _{13} \cos 2 \omega _{12} \cos\omega
   _{23}+\sin 2 \omega _{12} \sin\omega _{23} \left(\cos
   \omega _{13}+3 \cos 3 \omega _{13}\right)\right]\big\}\\
   &\quad+16 \sin\eta _1 \sin 2 \eta _2 \cos \eta _1 \sin 2
   \omega _{23} \cos2 \omega _{12} \cos2 \omega
   _{13}\bigg\}+4 \sin 2 \omega _{12} \sin \omega _{13} \big\{4
   \cos 2 \eta _1 \cos 2 \eta _2 \cos ^2\omega
   _{13}\\
   &\quad-\sin ^2\eta _1 \sin 2 \eta _2 \cos \eta
   _1 \left[4 \cos ^2\omega _{13}+\left(1-3 \cos2 \omega
   _{13}\right) \cos 2 \omega _{23}\right]\big\}\Bigg\},
   \end{split}
\end{equation}
\begin{equation}
\begin{split}
&I_1=\frac{(-1)^{k_1+k_2}}{8}  \sin \eta _1 \cos \omega _{13}
   \bigg\{4 \sqrt{2} \cos ^3\eta _1 \sin2
   \omega _{12} \sin\omega _{23} \cos ^2\omega
   _{13}-\sin ^3\eta _1\sin 2 \omega _{12} \sin 2 \omega
   _{13}  
\\
&\quad  +2 \sin \eta _1 \cos ^2\eta _1 \left[\sin2
   \omega _{12} \sin 2 \omega _{13} \left(2-\cos 2 \omega
   _{23}\right)-2 \sin 2 \omega _{23} \cos 2 \omega _{12}
   \cos\omega _{13}\right]\\
 &\quad    +\sqrt{2} \sin ^2\eta _1 \cos\eta _1 \left[\sin
   2 \omega _{12} \sin \omega _{23} \left(1-3 \cos 2 \omega
   _{13}\right)-4 \sin\omega _{13} \cos 2 \omega _{12} \cos
   \omega _{23}\right]\bigg\},
\end{split}
\end{equation}
\begin{equation}
\begin{split}
&I_2=\frac{ (-1)^{k_1+k_3} }{16}\Bigg\{2 \sin ^2\eta _1 \big\{4 \cos ^2\eta
   _1 \sin 2 \omega _{23} \left(\sin ^2\omega _{12}\sin
   ^2\omega _{13}+\cos2 \omega _{13} \cos ^2\omega
   _{12}\right)\\
&\quad  -\sin 2 \omega _{12}\sin \omega _{13}
   \left[\left(5 \cos 2 \eta _1+3\right) \cos ^2\omega _{13}+4 \cos
   ^2\eta _1 \sin ^2\omega _{13} \cos2 \omega
   _{23}\right]\big\}\\
&\quad   +\sqrt{2} \sin \eta _1\cos\eta _1
   \cos \omega _{13}
   \big\{\sin2 \omega _{12} \sin\omega
   _{23} \left[\cos2 \eta _1 \left(3-5 \cos 2 \omega
   _{13}\right)+2 \cos ^2\omega _{13}\right]\\
 &\quad -8 \sin\omega
   _{13} \cos\omega _{23} \left(\cos 2 \eta _1 \cos 2
   \omega _{12}+\cos ^2\eta _1\right)\big\}\Bigg\}.
   \end{split}
\end{equation}
\begin{figure}[tbh]
\center
\includegraphics[width=0.45\textwidth]{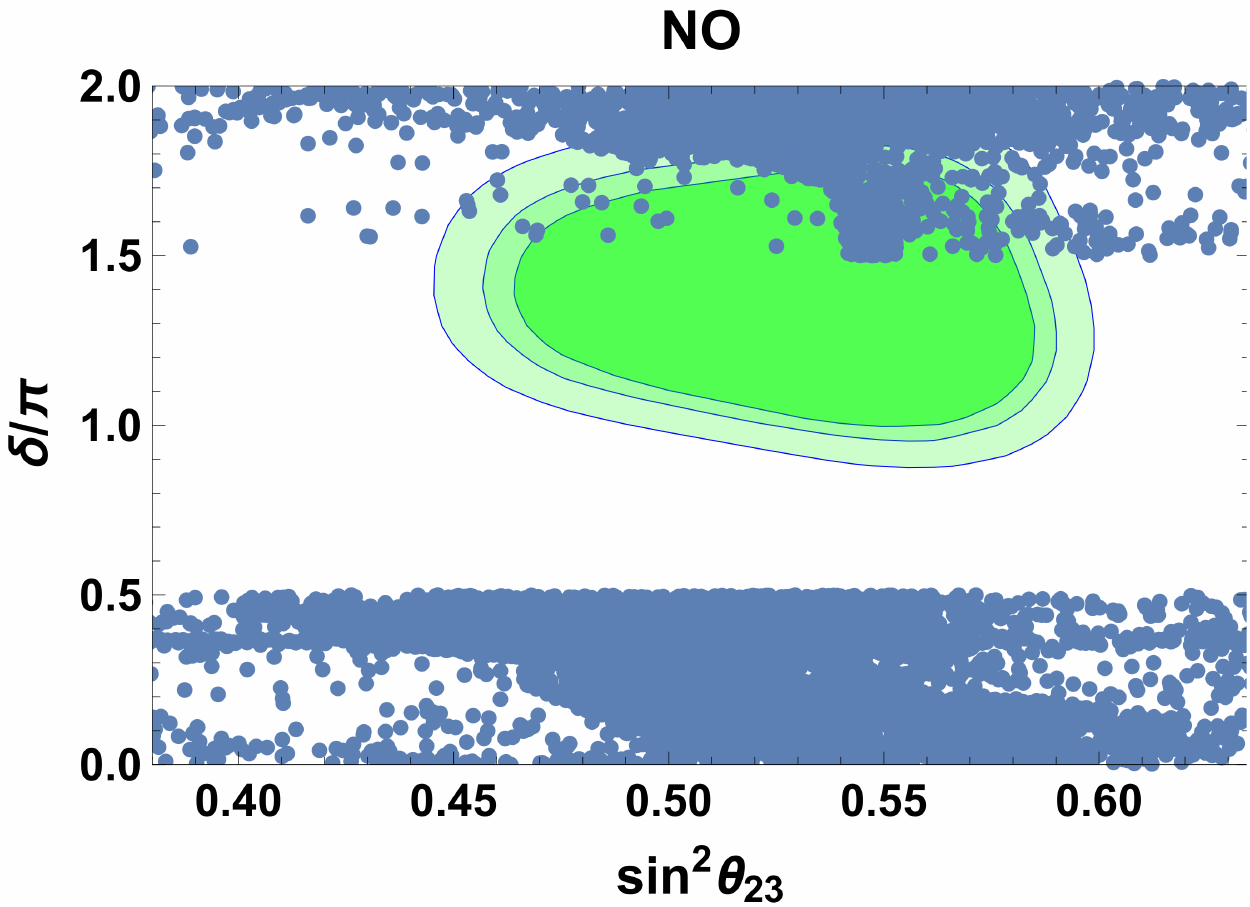} %
\includegraphics[width=0.45\textwidth]{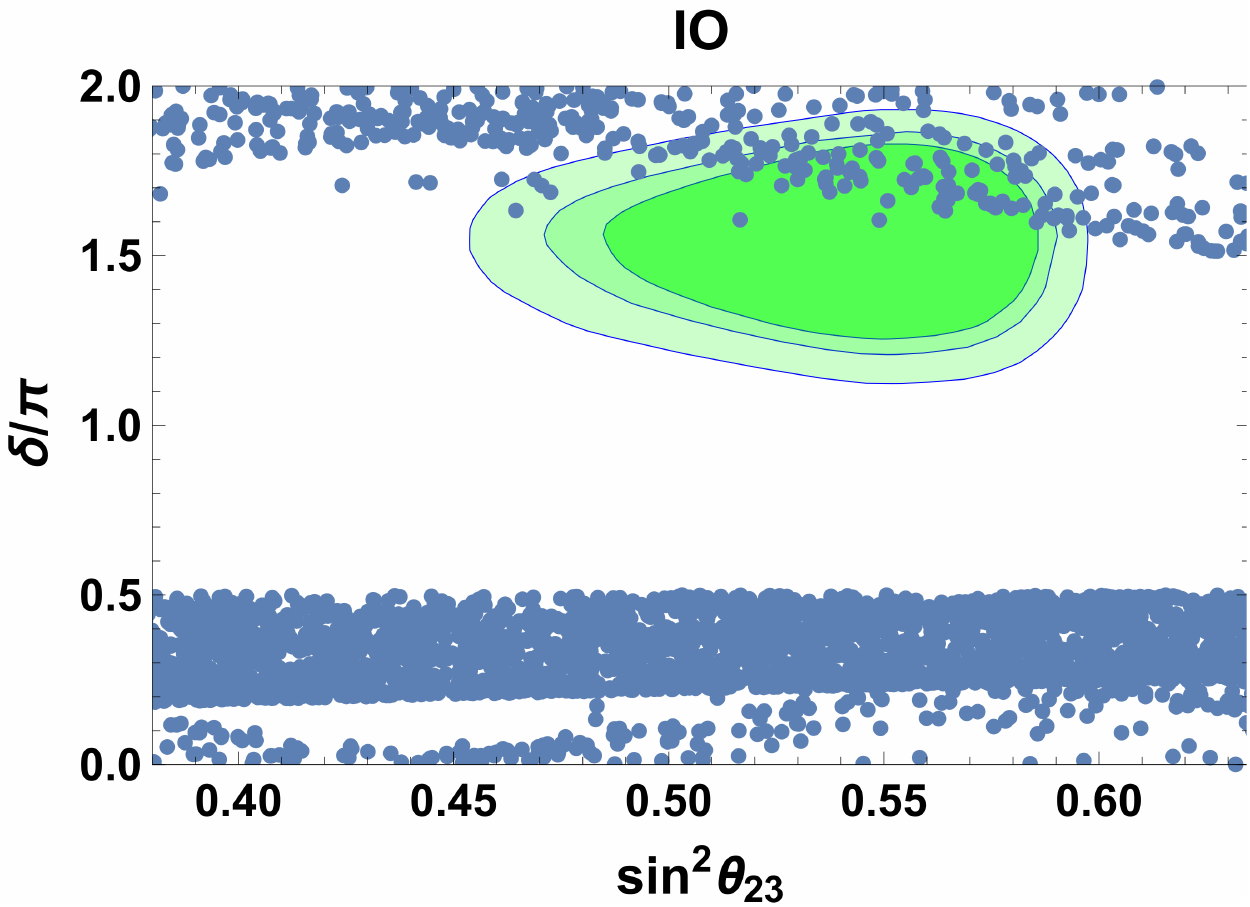}
\caption{Allowed values for the leptonic mixing parameters for Normal Ordering (NO)  and Inverted Ordering (IO). All points lie within the $2\sigma$ regions of the neutrino oscillation parameters. For comparison, the green regions correspond to the generic 90, 95 and 99\% C.L. regions from the Global fit in\cite{deSalas:2017kay}. }
\label{Correlation}
\end{figure}

In the above expressions, the angle $\omega_{12}$ can be eliminated using Eq.(\ref{pr}). Thus, the lepton mixing parameters depend on three angles $\omega_{13}$, $\omega_{23}$, $\psi$ (restricted in our analysis to satisfy Eq.(\ref{pr}) with real values for $\lambda_1$, $\lambda_2$ and $r$), two additional angles coming from charged lepton mixing and subject to $\eta_1\sim\eta_2\sim\lambda$ and three discrete variables $k_1$, $k_2$, $k_3$ entering in the Majorana phases.  

In Figure (\ref{Correlation}) we show the allowed values for the leptonic Dirac CP violating phase $\delta$ versus the atmospheric mixing parameter $\sin^2\theta_{23}$, for both normal and inverted neutrino mass orderings. These values were generated by randomly varying the model parameters $\omega_{13}$, $\omega_{23}$, $\psi$, $|\eta_1|$ and $|\eta_2|$  within a range that covers reactor and solar mixing angles inside the $2\sigma$ experimentally allowed range. In particular, we varied $|\eta_1|$ and $|\eta_2|$ in the range $[0.5\lambda,3\lambda]$. Furthermore, the light active neutrino mass scale was randomly varied in the range $10^{-4}\mathrm{eV}<m_0<1\mathrm{eV}$, consistent with $2\sigma$ allowed values for the neutrino mass squared splittings.

To close this section we note that, in contrast to the Left-right symmetric models of Refs.~\cite{Gomez-Izquierdo:2017rxi,Garces:2018nar}, where  the $\mu-\tau$ symmetry is broken softly, our departure from $\mu-\tau$ symmetry is induced by the mixing in the charged lepton sector, parameterized by the $\eta_1$ and $\eta_2$ angles, assumed to be of the same order as the Cabibbo angle $\lambda$.

\section{Neutrinoless double beta decay}
\label{sec:neutr-double-beta}

In this section we present the model predictions for neutrinoless double beta ($0\nu\beta\beta$) decay. The effective Majorana neutrino mass parameter is
\begin{equation}
\left\vert m_{\beta\beta}\right\vert = \left\vert
\sum_{i=1}^{3}m^{\nu}_{i}U_{ei}^{2}\right\vert=\left\vert m^{\nu}_{1}c^{2}_{12}c^{2}_{13}+m^{\nu}_{2}s^{2}_{12}c^{2}_{13}e^{-2i\phi_{12}}+m^{\nu}_{3}s^{2}_{13}e^{-2i\phi_{13}}\right\vert,
\label{meef} 
\end{equation}
where $m_{\nu_{i}}$ are the light active neutrino masses and $U_{ei}^{2}$ are the squared lepton mixing matrix elements, respectively. 
\begin{figure}[tbh]
\centering
\includegraphics[width=0.5\textwidth]{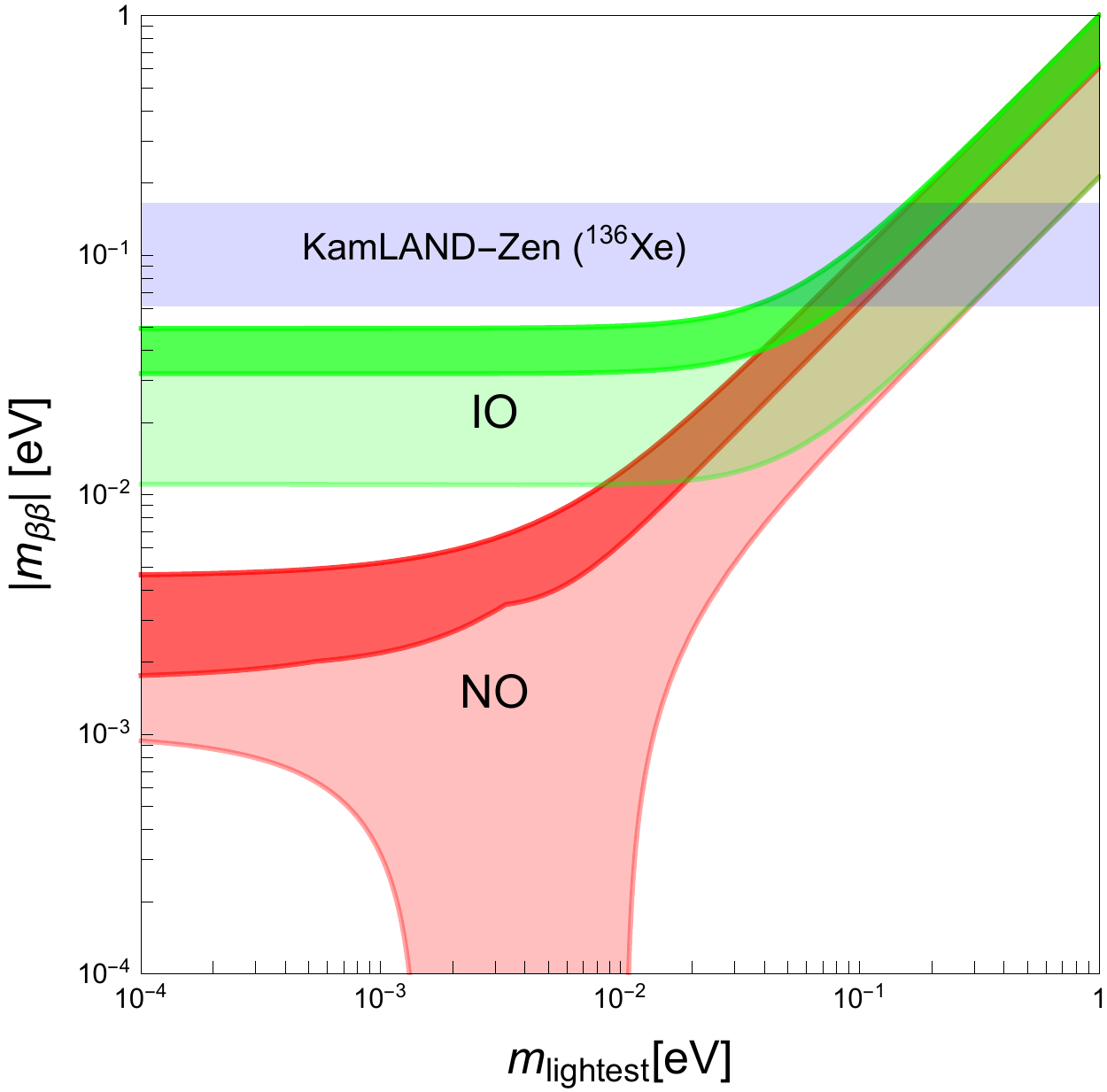}
\caption{Effective Majorana neutrino mass parameter $|m_{\beta\beta}|$ as a function of the lightest active neutrino mass $m_{\text{lightest}}$ for normal and inverted neutrino mass orderings. The dark regions correspond to points satisfying $2\sigma$ constraints for all mixing parameters. The current sensitivity of the KamLAND-Zen experiment is indicated in purple.}
\label{mee}
\end{figure}
The current experimental sensitivity on the Majorana neutrino mass parameter is illustrated by the horizontal band in Fig.~(\ref{mee}) and comes from the KamLAND-Zen limit on the ${}^{136}\mathrm{Xe}$ $0\nu\beta\beta$ decay half-life $T^{0\nu\beta\beta}_{1/2}({}^{136}\mathrm{Xe})\geq 1.07\times 10^{26}$ yr \cite{KamLAND-Zen:2016pfg}, which translates into a corresponding upper bound on $|m_{\beta\beta}|\leq (61-165)$ meV at 90\% C.L. as indicated by the horizontal band in Fig.~(\ref{mee}). For those of other experiments see Ref.~\cite{GERDA:2018,MAJORANA:2008,CUORE:2018,EXO-2018,Arnold:2016bed}.  
The ``expected'' regions for the effective Majorana neutrino mass parameter $|m_{\beta\beta}|$ consistent with the constraints from the current neutrino oscillation data at the $2\sigma$ level are indicated by the other broad shaded bands. There are two cases, corresponding to normal and inverted neutrino mass orderings. These are generic, arising only by imposing current oscillation data. In contrast, the thinner (darker) bands include also the model predictions described in the previous section. These regions are obtained from our generated model points by imposing current neutrino oscillation constraints at the $2\sigma$ level. 

One sees that our ``predicted'' ranges for the effective Majorana neutrino mass parameter have lower bounds in both cases, of normal and inverted mass orderings, indicating that a complete destructive interference amongst the three light neutrinos is always prevented in our model. These lower bounds for the \nbb amplitude are general predictions of the present model, and can easily be understood. 

In fact,
as mentioned above, the structure of our $\mu-\tau$ symmetric neutrino mass matrix is a particular case of that in Ref.~\cite{Grimus:2003yn}. Comparing with the results of Ref.~\cite{Chen:2015siy} one sees that, indeed, the possible destructive interference amongst the three light neutrinos is prevented (as $\lim_{\eta_1\rightarrow 0} I_{1,2}=0$), thus explaining the absolute lower bound we obtain.

The experimental sensitivity of \nbb searches is expected to improve in the near future. For our model, the predicted \nbb decay rates may be tested by the next-generation bolometric CUORE experiment \cite{Alduino:2017pni}, as well as the next-to-next-generation ton-scale $0\nu \beta \beta $-decay experiments \cite{KamLAND-Zen:2016pfg,Albert:2017owj,Abt:2004yk,Gilliss:2018lke}.

\section{Quark masses and mixings}

\label{quarkmassesandmixing}

In this section, we illustrate how the model is capable of reproducing the correct masses and mixings in the quark sector. From the quark Yukawa interactions, we find that the up-type mass matrix in the basis $(\overline{u}_{1L},\overline{u}_{2L},\overline{u}_{3L},\overline{T}_{1L},\overline{T}_{2L},\overline{T}_{3L})$ versus $(u_{1R},u_{2R},u_{3R},T_{1R},T_{2R},T_{3R})$ is given by:
\begin{equation}
M_{U}=\left( 
\begin{array}{ccc}
0_{2\times 2} & 0_{2\times 1} & x\frac{v_{L}}{\sqrt{2}} \\ 
0_{1\times 2} & \alpha _{33}\frac{v}{\sqrt{2}} & 0_{1\times 2} \\ 
x^{T}\frac{v_{R}}{\sqrt{2}} & 0_{2\times 1} & M_{T}%
\end{array}%
\right) ,\hspace{1cm}x=\left( 
\begin{array}{cc}
x_{11}\lambda ^{2} & x_{12}\lambda \\ 
0 & x_{22}%
\end{array}%
\right) ,\hspace{1cm}M_{T}=\left( 
\begin{array}{cc}
m_{T_{1}} & 0 \\ 
0 & m_{T_{2}}%
\end{array}%
\right) ,\hspace{1cm}m_{t}=\alpha \frac{v}{\sqrt{2}},
\end{equation}
while the down type quark mass matrix is given as:
\begin{equation}
M_{D}=\left( 
\begin{array}{cc}
0_{3\times 3} & y\frac{v_{L}}{\sqrt{2}} \\ 
y^{T}\frac{v_{R}}{\sqrt{2}} & M_{B}%
\end{array}%
\right) ,\hspace{1cm}y=\left( 
\begin{array}{ccc}
y_{11}\lambda ^{2} & 0 & y_{13}\lambda ^{3} \\ 
0 & y_{22}\lambda & y_{23}\lambda ^{2} \\ 
0 & 0 & y_{33}%
\end{array}%
\right) \allowbreak ,\hspace{1cm}M_{B}=\left( 
\begin{array}{ccc}
m_{B_{1}} & 0 & 0 \\ 
0 & m_{B_{2}} & 0 \\ 
0 & 0 & m_{B_{3}}%
\end{array}%
\right) ,\allowbreak \allowbreak
\end{equation}%
expressed in the basis $(\overline{d}_{1L},\overline{d}_{2L},\overline{d}_{3L},\overline{B}_{1L},\overline{B}_{2L},\overline{B}_{3L})$-$(d_{1R},d_{2R},d_{3R},B_{1R},B_{2R},B_{3R})$.
Assuming the exotic quark masses to be sufficiently larger than $v_{L}$ and $v_{R}$, it follows that the SM quarks lighter than the top quark all get their masses from a Universal seesaw mechanism mediated by the three  exotic up-type and down-type quarks $U_{i}$ and $D_{i}$ ($i=1,2,3$). It is worth mentioning that the top quark does not mix with the remaining up-type quarks. As a result, the SM quark mass matrices take the form:
\begin{eqnarray}
\widetilde{M}_{U} &=&\left( 
\begin{array}{cc}
\frac{v_{L}v_{R}}{2}xM_{T}^{-1}x^{T} & 0_{2\times 1} \\ 
0_{1\times 2} & m_{t}%
\end{array}%
\right) =\left( 
\begin{array}{ccc}
\left( x_{12}^{2}+\frac{m_{T_{2}}}{m_{T_{1}}}x_{11}^{2}\lambda ^{2}\right)
\lambda ^{2}\frac{v_{L}v_{R}}{2m_{T_{2}}} & x_{12}x_{22}\lambda \frac{%
v_{L}v_{R}}{2m_{T_{2}}} & 0 \\ 
x_{12}x_{22}\lambda \frac{v_{L}v_{R}}{2m_{T_{2}}} & x_{22}^{2}\frac{%
v_{L}v_{R}}{2m_{T_{2}}} & 0 \\ 
0 & 0 & m_{t}%
\end{array}%
\right)  \notag \\
\allowbreak \allowbreak &=&\left( 
\begin{array}{ccc}
\left( \frac{a_{12}^{2}}{a_{22}}+\kappa \lambda ^{2}\right) \lambda ^{6} & 
a_{12}\lambda ^{5} & 0 \\ 
a_{12}\lambda ^{5} & a_{22}\lambda ^{4} & 0 \\ 
0 & 0 & \alpha%
\end{array}%
\right) \allowbreak \allowbreak \frac{v}{\sqrt{2}},
\label{eq:MU-mass}
\end{eqnarray}
\begin{eqnarray}
\widetilde{M}_{D} &=&\frac{v_{L}v_{R}}{2}yM_{B}^{-1}y^{T}=\left( 
\begin{array}{ccc}
\left( \frac{m_{B_{3}}}{m_{B_{1}}}y_{11}^{2}+y_{13}^{2}\lambda ^{2}\right)
\lambda ^{4} & y_{13}y_{23}\lambda ^{5} & y_{13}y_{33}\lambda ^{3} \\ 
y_{13}y_{23}\lambda ^{5} & \left( \frac{m_{B_{3}}}{m_{B_{2}}}%
y_{22}^{2}+y_{23}^{2}\lambda ^{2}\right) \lambda ^{2} & y_{23}y_{33}\lambda
^{2} \\ 
y_{13}y_{33}\lambda ^{3} & y_{23}y_{33}\lambda ^{2} & y_{33}^{2}%
\end{array}%
\right) \allowbreak \frac{v_{L}v_{R}}{2m_{B_{3}}}  \notag \\
&=&\left( 
\begin{array}{ccc}
b_{11}\lambda ^{7} & b_{12}\lambda ^{8} & b_{13}\lambda ^{6} \\ 
b_{12}\lambda ^{8} & b_{22}\lambda ^{5} & b_{23}\lambda ^{5} \\ 
b_{13}\lambda ^{6} & b_{23}\lambda ^{5} & b_{33}\lambda ^{3}%
\end{array}%
\right) \allowbreak \frac{v}{\sqrt{2}},
\label{eq:MD-mass}
\end{eqnarray}
where we have set $v_{L}=\lambda ^{4}\frac{\sqrt{2}v}{v_{R}} m_{T_{2}}=\lambda ^{3}\frac{\sqrt{2}v}{v_{R}}m_{B_{2}}$. Let us note that in our model, the dominant contribution to the Cabbibo mixing arises from the up-type quark sector, whereas the down-type quark sector contributes to the remaining CKM mixing angles. In order to recover the low energy quark flavor data, we assume that all dimensionless parameters of the SM quark mass matrices are real, except for $b_{13}$, taken to be complex. 

Starting from the following benchmark point:
\begin{eqnarray}
\label{eq:Quark-benchmark-point}
a_{12} &\simeq &-1.375,\hspace{1cm}a_{22}\simeq 1.599,\hspace{1cm}\kappa
\simeq 1.724,\hspace{1cm}\alpha \simeq 0.989,\hspace{1cm}b_{11}\simeq 0.595,%
\hspace{1cm}b_{12}\simeq 0.847,  \notag \\
b_{22} &\simeq &0.640,\hspace{1cm}\left\vert b_{13}\right\vert \simeq 1.168,%
\hspace{1cm}\arg \left( b_{13}\right) \simeq -158.2^{\circ },\hspace{1cm}%
b_{23}\simeq 1.104,\hspace{1cm}b_{33}\simeq 1.414
\end{eqnarray}
\begin{table}[tbh]
\begin{center}
\begin{tabular}{c|l|l}
\hline\hline
Observable & Model value & Experimental value \\ \hline
$m_{u}(\mathrm{MeV})$ & \quad $1.92$ & \quad $1.45_{-0.45}^{+0.56}$ \\ \hline
$m_{c}(\mathrm{MeV})$ & \quad $744$ & \quad $635\pm 86$ \\ \hline
$m_{t}(\mathrm{GeV})$ & \quad $172.1$ & \quad $172.1\pm 0.6\pm 0.9$ \\ \hline
$m_{d}(\mathrm{MeV})$ & \quad $2.82$ & \quad $2.9_{-0.4}^{+0.5}$ \\ \hline
$m_{s}(\mathrm{MeV})$ & \quad $60.3$ & \quad $57.7_{-15.7}^{+16.8}$ \\ \hline
$m_{b}(\mathrm{GeV})$ & \quad $2.82$ & \quad $2.82_{-0.04}^{+0.09}$ \\ \hline
$\sin \theta _{12}$ & \quad $0.225$ & \quad $0.22536\pm 0.00061$ \\ \hline
$\sin \theta _{23}$ & \quad $0.0414$ & \quad $0.0414\pm 0.0012$ \\ \hline
$\sin \theta _{13}$ & \quad $0.00355$ & \quad $0.00355\pm 0.00015$ \\ \hline
$J$ & \quad $3.08\times 10^{-5}$ & \quad $2.96_{-0.16}^{+0.20}\times 10^{-5}$
\\ \hline\hline
\end{tabular}%
\end{center}
\caption{Model and experimental values of the quark masses and CKM
parameters.}
\label{Tab}
\end{table}
one can check that the resulting values for the physical quark mass spectrum \cite{Bora:2012tx,Xing:2007fb}, mixing angles and Jarlskog invariant \cite{Olive:2016xmw} are indeed consistent with the experimental data, as shown in Table \ref{Tab}. This establishes the viability of our model also for the quark sector. 
Note that the dimensionless parameters of the benchmark point (\ref{eq:Quark-benchmark-point}) are all $\sim O(1)$ in absolute value. This means that our model reproduces the quark mass and mixing hierarchy by its symmetries resulting in certain distribution of the powers of $\lambda$ among the entries of the mass matrices 
(\ref{eq:MU-mass}), (\ref{eq:MD-mass}).
\\[.5cm]

\section{Features of the model}

We now sum up the main theoretical features of our model.
\begin{enumerate}
\item Only the top quark and the gauge singlet Majorana neutrinos $\Omega_i$ ($i=1,2,3$) acquire masses from renormalizable Yukawa interactions. The exotic charged fermions all have bare tree level mass terms.  
\item The masses for the SM charged fermions lighter than the top quark arise from a Universal Seesaw mechanism mediated by charged exotic fermions. The quark mixing angles and the hierarchy between quark masses arise from the spontaneous breaking of the $Z_6\otimes Z_{12}$ discrete group. 
\item The Cabibbo mixing arises from the up-type quark sector, whereas the down-type quark sector induces the remaining CKM mixing angles. On the other hand,  
the leptonic mixing parameters receive their dominant contributions from the light active neutrino mass matrix, whereas the charged lepton mass matrix provides Cabibbo-sized corrections to these parameters.
\item The masses for the light active neutrinos emerge from a one loop level inverse seesaw mechanism, whose radiative nature is guaranteed by the spontaneously broken $Z_{4}$ and $Z_{12}$ symmetries, with $Z_{12}$ spontaneously broken down to a preserved $Z_2$ symmetry.
\item The mass terms for the gauge singlet sterile neutrinos $S_i$ ($i=1,2,3$) are generated from at one loop level, mediated by the real and imaginary components of the electrically neutral gauge singlet scalar $\varphi$ as well as by the gauge singlet Majorana neutrinos $\Omega_i$ ($i=1,2,3$). These mass terms break lepton number by two units, triggering the one loop level inverse seesaw mechanism responsible for the light active neutrino masses.
\end{enumerate}

\section{Discussion and Conclusions}
\label{sec:conclusions}
In summary, we have built a viable extension of the left-right symmetric electroweak extension of the \sm capable of explaining the current pattern of SM fermion masses and mixings. Our model is based on the $\Delta(27)$ discrete symmetry, supplemented by the $Z_4\otimes Z_6\otimes Z_{12}$ discrete family group. In our model, the masses of the light active neutrinos emerge from a one loop level inverse seesaw mechanism, whereas the masses of the \sm charged fermions lighter than the top quark are produced by a Universal Seesaw mechanism. Of the Standard Model fermions only the top quark acquires mass through a tree level renormalizable Yukawa interaction. In our model the Cabibbo mixing arises from the up-type quark sector whereas the down-type quark sector contributes to the other CKM mixing angles. On the other hand, the leptonic mixing parameters receive their dominant contributions from the light active neutrino mass matrix, whereas the SM charged lepton mass matrix provide Cabibbo sized corrections. The observed hierarchy of SM charged fermion masses and mixing angles is caused by the spontaneous breaking of the $\Delta \left( 27\right)\otimes Z_{6}\otimes Z_{12}$ discrete flavor group, whereas the radiative nature of the inverse seessaw mechanism is guaranteed by spontaneously broken $Z_{4}$ and $Z_{12}$ symmetries, having $Z_{12}$ spontaneously broken down to a preserved $Z_2$ symmetry. Our model features a generalized $\mu-\tau$ symmetry and predicts a restricted range of neutrino oscillations parameters, with the neutrinoless double beta decay amplitude lying at the upper ranges associated to normal and inverted neutrino mass ordering. 

Notice also that our low-scale left-right symmetric radiative seesaw scheme not only accounts for the light neutrino masses and mixings that lead to oscillations and $0\nu \beta \beta $-decay, but can also lead to signatures that can make it testable at collider experiments such as the LHC. 
For example, the heavy quasi Dirac neutrinos can be produced in pairs at the LHC, via a Drell-Yan mechanism mediated by a heavy non Standard Model neutral gauge boson $Z^{\prime }$.
These heavy quasi Dirac neutrinos can decay into a Standard Model charged lepton and $W$ gauge boson, due to their mixings with the light active neutrinos. 
Thus, the observation of an excess of events in the dilepton final states with respect to the SM background, would be a signal supporting this model at the LHC.
Moreover, lepton flavor violation is expected in these decays, even if suppressed at low energies~\cite{Das:2012ii,Deppisch:2013cya}.
A detailed study of the collider phenomenology of this model is beyond the scope of the present paper and is left for future studies.

\section*{Acknowledgments}

Work supported by the Spanish grants SEV-2014-0398 and FPA2017-85216-P (AEI/FEDER, UE), PROMETEO/2018/165 (Generalitat Valenciana) and the Spanish Red Consolider MultiDark FPA201790566REDC; by the Chilean grants Fondecyt No. 1170803, No. 1150792 and CONICYT PIA/Basal FB0821 and ACT1406; and the Mexican C\'atedras CONACYT project 749. AECH is very grateful to Institut de F\'{i}sica Corpuscular (IFIC), where part of this work was done, for hospitality and University of Southampton for hospitality during the completion of this work. 

\bibliographystyle{utphys}
\bibliography{bibliography}

\providecommand{\href}[2]{#2}\begingroup\raggedright\begin{thebibliography}{10}

\bibitem{deSalas:2017kay}
P.~F. de~Salas {\em et~al.}, ``{Status of neutrino oscillations 2018: 3$\sigma$
  hint for normal mass ordering and improved CP sensitivity},''
  \href{http://dx.doi.org/10.1016/j.physletb.2018.06.019}{{\em Phys. Lett.}
  {\bfseries B782} (2018) 633--640},
  \href{http://arxiv.org/abs/1708.01186}{{\ttfamily arXiv:1708.01186
  [hep-ph]}}.
\url{http://globalfit.astroparticles.es/}.

\bibitem{Pati:1974yy}
J.~C. Pati and A.~Salam, ``{Lepton Number as the Fourth Color},'' {\em Phys.
  Rev.} {\bfseries D10} (1974) 275--289.
[Erratum: Phys. Rev.D11,703(1975)].

\bibitem{Mohapatra:1974gc}
R.~Mohapatra and J.~C. Pati, ``{A Natural Left-Right Symmetry},'' {\em
  Phys.Rev.} {\bfseries D11} (1975) 2558.

\bibitem{Ishimori:2010au}
H.~Ishimori, T.~Kobayashi, H.~Ohki, Y.~Shimizu, H.~Okada, and M.~Tanimoto,
  ``{Non-Abelian Discrete Symmetries in Particle Physics},''
  \href{http://dx.doi.org/10.1143/PTPS.183.1}{{\em Prog. Theor. Phys. Suppl.}
  {\bfseries 183} (2010) 1--163},
\href{http://arxiv.org/abs/1003.3552}{{\ttfamily arXiv:1003.3552 [hep-th]}}.

\bibitem{Morisi:2012fg}
S.~Morisi and J.~W.~F. Valle, ``{Neutrino masses and mixing: a flavour symmetry
  roadmap},'' \href{http://dx.doi.org/10.1002/prop.201200125}{{\em
  Fortsch.Phys.} {\bfseries 61} (2013) 466--492},
  \href{http://arxiv.org/abs/1206.6678}{{\ttfamily arXiv:1206.6678 [hep-ph]}}.

\bibitem{King:2014nza}
S.~F. King, A.~Merle, S.~Morisi, Y.~Shimizu, and M.~Tanimoto, ``{Neutrino Mass
  and Mixing: from Theory to Experiment},''
  \href{http://dx.doi.org/10.1088/1367-2630/16/4/045018}{{\em New J. Phys.}
  {\bfseries 16} (2014) 045018},
\href{http://arxiv.org/abs/1402.4271}{{\ttfamily arXiv:1402.4271 [hep-ph]}}.

\bibitem{CarcamoHernandez:2017owh}
A.~E. Carcamo~Hernandez, S.~Kovalenko, J.~W.~F. Valle, and C.~A.
  Vaquera-Araujo, ``{Predictive Pati-Salam theory of fermion masses and
  mixing},'' \href{http://dx.doi.org/10.1007/JHEP07(2017)118}{{\em JHEP}
  {\bfseries 07} (2017) 118},
\href{http://arxiv.org/abs/1705.06320}{{\ttfamily arXiv:1705.06320 [hep-ph]}}.

\bibitem{Mohapatra:1986bd}
R.~N. Mohapatra and J.~W.~F. Valle, ``{Neutrino mass and baryon-number
  nonconservation in superstring models},''
  \href{http://dx.doi.org/10.1103/PhysRevD.34.1642}{{\em Phys. Rev.} {\bfseries
  D34} (1986) 1642}.

\bibitem{GonzalezGarcia:1988rw}
M.~Gonzalez-Garcia and J.~W.~F. Valle, ``{Fast Decaying Neutrinos and
  Observable Flavor Violation in a New Class of Majoron Models},''
  \href{http://dx.doi.org/10.1016/0370-2693(89)91131-3}{{\em Phys.Lett.}
  {\bfseries B216} (1989) 360}.

\bibitem{Akhmedov:1995vm}
E.~K. Akhmedov {\em et~al.}, ``{Dynamical left-right symmetry breaking},''
  \href{http://dx.doi.org/10.1103/PhysRevD.53.2752}{{\em Phys.Rev.} {\bfseries
  D53} (1996) 2752--2780},
  \href{http://arxiv.org/abs/hep-ph/9509255}{{\ttfamily arXiv:hep-ph/9509255
  [hep-ph]}}.

\bibitem{Akhmedov:1995ip}
E.~K. Akhmedov {\em et~al.}, ``{Left-right symmetry breaking in NJL
  approach},'' \href{http://dx.doi.org/10.1016/0370-2693(95)01504-3}{{\em
  Phys.Lett.} {\bfseries B368} (1996) 270--280},
  \href{http://arxiv.org/abs/hep-ph/9507275}{{\ttfamily arXiv:hep-ph/9507275
  [hep-ph]}}.

\bibitem{Malinsky:2005bi}
M.~Malinsky, J.~Romao, and J.~W.~F. Valle, ``{Novel supersymmetric SO(10)
  seesaw mechanism},''
  \href{http://dx.doi.org/10.1103/PhysRevLett.95.161801}{{\em Phys.Rev.Lett.}
  {\bfseries 95} (2005) 161801},
  \href{http://arxiv.org/abs/hep-ph/0506296}{{\ttfamily arXiv:hep-ph/0506296
  [hep-ph]}}.

\bibitem{Bazzocchi:2009kc}
F.~Bazzocchi, D.~G. Cerdeno, C.~Munoz, and J.~W.~F. Valle, ``{Calculable
  inverse-seesaw neutrino masses in supersymmetry},''
  \href{http://dx.doi.org/10.1103/PhysRevD.81.051701}{{\em Phys. Rev.}
  {\bfseries D81} (2010) 051701},
\href{http://arxiv.org/abs/0907.1262}{{\ttfamily arXiv:0907.1262 [hep-ph]}}.

\bibitem{CarcamoHernandez:2017kra}
A.~E. Carcamo~Hernandez and H.~N. Long, ``{A highly predictive $A_{4}$ flavour
  3-3-1 model with radiative inverse seesaw mechanism},''
  \href{http://dx.doi.org/10.1088/1361-6471/aaace7}{{\em J. Phys.} {\bfseries
  G45} no.~4, (2018) 045001},
\href{http://arxiv.org/abs/1705.05246}{{\ttfamily arXiv:1705.05246 [hep-ph]}}.

\bibitem{Davidson:1987mh}
A.~Davidson and K.~C. Wali, ``{Universal Seesaw Mechanism?},''
\href{http://dx.doi.org/10.1103/PhysRevLett.59.393}{{\em Phys. Rev. Lett.}
  {\bfseries 59} (1987) 393}.

\bibitem{Berezhiani:1991ds}
Z.~G. Berezhiani and R.~Rattazzi, ``{Universal seesaw and radiative quark mass
  hierarchy},''
\href{http://dx.doi.org/10.1016/0370-2693(92)91851-Y}{{\em Phys. Lett.}
  {\bfseries B279} (1992) 124--130}.

\bibitem{Sogami:1991yq}
I.~S. Sogami and T.~Shinohara, ``{Universal seesaw mechanism for quarks and
  leptons},''
\href{http://dx.doi.org/10.1143/PTP.86.1031}{{\em Prog. Theor. Phys.}
  {\bfseries 86} (1991) 1031--1052}.

\bibitem{Gu:2010zv}
P.-H. Gu and M.~Lindner, ``{Universal Seesaw from Left-Right and Peccei-Quinn
  Symmetry Breaking},''
  \href{http://dx.doi.org/10.1016/j.physletb.2011.02.042}{{\em Phys. Lett.}
  {\bfseries B698} (2011) 40--43},
\href{http://arxiv.org/abs/1010.4635}{{\ttfamily arXiv:1010.4635 [hep-ph]}}.

\bibitem{Alvarado:2012xi}
C.~Alvarado, R.~Martinez, and F.~Ochoa, ``{Quark mass hierarchy in 3-3-1
  models},'' \href{http://dx.doi.org/10.1103/PhysRevD.86.025027}{{\em Phys.
  Rev.} {\bfseries D86} (2012) 025027},
\href{http://arxiv.org/abs/1207.0014}{{\ttfamily arXiv:1207.0014 [hep-ph]}}.

\bibitem{Hernandez:2013mcf}
A.~E. Carcamo~Hernandez, R.~Martinez, and F.~Ochoa, ``{Radiative seesaw-type
  mechanism of quark masses in $SU(3)_C \otimes SU(3)_L \otimes U(1)_X$},''
  \href{http://dx.doi.org/10.1103/PhysRevD.87.075009}{{\em Phys. Rev.}
  {\bfseries D87} no.~7, (2013) 075009},
\href{http://arxiv.org/abs/1302.1757}{{\ttfamily arXiv:1302.1757 [hep-ph]}}.

\bibitem{Kawasaki:2013apa}
R.~Kawasaki, T.~Morozumi, and H.~Umeeda, ``{Quark sector CP violation of the
  universal seesaw model},''
  \href{http://dx.doi.org/10.1103/PhysRevD.88.033019}{{\em Phys. Rev.}
  {\bfseries D88} (2013) 033019},
\href{http://arxiv.org/abs/1306.5080}{{\ttfamily arXiv:1306.5080 [hep-ph]}}.

\bibitem{Mohapatra:2014qva}
R.~N. Mohapatra and Y.~Zhang, ``{TeV Scale Universal Seesaw, Vacuum Stability
  and Heavy Higgs},'' \href{http://dx.doi.org/10.1007/JHEP06(2014)072}{{\em
  JHEP} {\bfseries 06} (2014) 072},
\href{http://arxiv.org/abs/1401.6701}{{\ttfamily arXiv:1401.6701 [hep-ph]}}.

\bibitem{Dev:2015vjd}
P.~S.~B. Dev, R.~N. Mohapatra, and Y.~Zhang, ``{Quark Seesaw, Vectorlike
  Fermions and Diphoton Excess},''
  \href{http://dx.doi.org/10.1007/JHEP02(2016)186}{{\em JHEP} {\bfseries 02}
  (2016) 186},
\href{http://arxiv.org/abs/1512.08507}{{\ttfamily arXiv:1512.08507 [hep-ph]}}.

\bibitem{Borah:2017inr}
D.~Borah and S.~Patra, ``{Universal seesaw and $0\nu\beta\beta$ in new 3331
  left-right symmetric model},''
  \href{http://dx.doi.org/10.1016/j.physletb.2017.05.059}{{\em Phys. Lett.}
  {\bfseries B771} (2017) 318--326},
\href{http://arxiv.org/abs/1701.08675}{{\ttfamily arXiv:1701.08675 [hep-ph]}}.

\bibitem{Patra:2017gak}
A.~Patra and S.~K. Rai, ``{Lepton-specific universal seesaw model with
  left-right symmetry},''
  \href{http://dx.doi.org/10.1103/PhysRevD.98.015033}{{\em Phys. Rev.}
  {\bfseries D98} no.~1, (2018) 015033},
\href{http://arxiv.org/abs/1711.00627}{{\ttfamily arXiv:1711.00627 [hep-ph]}}.

\bibitem{King:2018fcg}
S.~F. King, ``{$ {R}_{K^{\left(*\right)}} $ and the origin of Yukawa
  couplings},'' \href{http://dx.doi.org/10.1007/JHEP09(2018)069}{{\em JHEP}
  {\bfseries 09} (2018) 069},
\href{http://arxiv.org/abs/1806.06780}{{\ttfamily arXiv:1806.06780 [hep-ph]}}.

\bibitem{Babu:2018vrl}
K.~S. Babu, R.~N. Mohapatra, and B.~Dutta, ``{A Theory of $R(D^*,D)$ Anomaly
  with Right-Handed Currents},''
\href{http://arxiv.org/abs/1811.04496}{{\ttfamily arXiv:1811.04496 [hep-ph]}}.

\bibitem{Branco:1983tn}
G.~Branco, J.-M. Gerard, and W.~Grimus, ``Geometrical t-violation,''
  \href{http://dx.doi.org/https://doi.org/10.1016/0370-2693(84)92024-0}{{\em
  Physics Letters B} {\bfseries 136} no.~5, (1984) 383 -- 386}.
  \url{http://www.sciencedirect.com/science/article/pii/0370269384920240}.

\bibitem{Bernal:2017xat}
N.~Bernal, A.~E. Carcamo~Hernandez, I.~de~Medeiros~Varzielas, and S.~Kovalenko,
  ``{Fermion masses and mixings and dark matter constraints in a model with
  radiative seesaw mechanism},''
  \href{http://dx.doi.org/10.1007/JHEP05(2018)053}{{\em JHEP} {\bfseries 05}
  (2018) 053},
\href{http://arxiv.org/abs/1712.02792}{{\ttfamily arXiv:1712.02792 [hep-ph]}}.

\bibitem{Bhattacharyya:2012pi}
G.~Bhattacharyya, I.~de~Medeiros~Varzielas, and P.~Leser, ``{A common origin of
  fermion mixing and geometrical CP violation, and its test through Higgs
  physics at the LHC},''
  \href{http://dx.doi.org/10.1103/PhysRevLett.109.241603}{{\em Phys. Rev.
  Lett.} {\bfseries 109} (2012) 241603},
\href{http://arxiv.org/abs/1210.0545}{{\ttfamily arXiv:1210.0545 [hep-ph]}}.

\bibitem{Aranda:2013gga}
A.~Aranda {\em et~al.}, ``{Dirac neutrinos from flavor symmetry},''
  \href{http://dx.doi.org/10.1103/PhysRevD.89.033001}{{\em Phys. Rev.}
  {\bfseries D89} no.~3, (2014) 033001},
  \href{http://arxiv.org/abs/1307.3553}{{\ttfamily arXiv:1307.3553 [hep-ph]}}.

\bibitem{CarcamoHernandez:2018djj}
A.~E. Cárcamo~Hernández, J.~C. Gómez-Izquierdo, S.~Kovalenko, and
  M.~Mondragón, ``{$\Delta \left( 27\right)$ flavor singlet-triplet Higgs
  model for fermion masses and mixings},''
\href{http://arxiv.org/abs/1810.01764}{{\ttfamily arXiv:1810.01764 [hep-ph]}}.

\bibitem{Bjorkeroth:2015uou}
F.~Björkeroth, F.~J. de~Anda, I.~de~Medeiros~Varzielas, and S.~F. King,
  ``{Towards a complete $\Delta(27) \times SO(10)$ SUSY GUT},''
  \href{http://dx.doi.org/10.1103/PhysRevD.94.016006}{{\em Phys. Rev.}
  {\bfseries D94} no.~1, (2016) 016006},
\href{http://arxiv.org/abs/1512.00850}{{\ttfamily arXiv:1512.00850 [hep-ph]}}.

\bibitem{deMedeirosVarzielas:2017sdv}
I.~de~Medeiros~Varzielas, G.~G. Ross, and J.~Talbert, ``{A Unified Model of
  Quarks and Leptons with a Universal Texture Zero},''
  \href{http://dx.doi.org/10.1007/JHEP03(2018)007}{{\em JHEP} {\bfseries 03}
  (2018) 007},
\href{http://arxiv.org/abs/1710.01741}{{\ttfamily arXiv:1710.01741 [hep-ph]}}.

\bibitem{deMedeirosVarzielas:2018vab}
I.~De~Medeiros~Varzielas, M.~L. López-Ibáñez, A.~Melis, and O.~Vives,
  ``{Controlled flavor violation in the MSSM from a unified $\Delta(27)$ flavor
  symmetry},''
\href{http://arxiv.org/abs/1807.00860}{{\ttfamily arXiv:1807.00860 [hep-ph]}}.

\bibitem{Chen:2015jta}
P.~Chen, G.-J. Ding, A.~D. Rojas, C.~A. Vaquera-Araujo, and J.~W.~F. Valle,
  ``{Warped flavor symmetry predictions for neutrino physics},''
  \href{http://dx.doi.org/10.1007/JHEP01(2016)007}{{\em JHEP} {\bfseries 01}
  (2016) 007},
\href{http://arxiv.org/abs/1509.06683}{{\ttfamily arXiv:1509.06683 [hep-ph]}}.

\bibitem{Vien:2016tmh}
V.~V. Vien, A.~E. Carcamo~Hernandez, and H.~N. Long, ``{The $\Delta(27)$ flavor
  3-3-1 model with neutral leptons},''
  \href{http://dx.doi.org/10.1016/j.nuclphysb.2016.10.010}{{\em Nucl. Phys.}
  {\bfseries B913} (2016) 792--814},
\href{http://arxiv.org/abs/1601.03300}{{\ttfamily arXiv:1601.03300 [hep-ph]}}.

\bibitem{Hernandez:2016eod}
A.~E. Carcamo~Hernandez, H.~N. Long, and V.~V. Vien, ``{A 3-3-1 model with
  right-handed neutrinos based on the $\varDelta \left( 27\right) $ family
  symmetry},'' \href{http://dx.doi.org/10.1140/epjc/s10052-016-4074-0}{{\em
  Eur. Phys. J.} {\bfseries C76} no.~5, (2016) 242},
\href{http://arxiv.org/abs/1601.05062}{{\ttfamily arXiv:1601.05062 [hep-ph]}}.

\bibitem{CarcamoHernandez:2018iel}
A.~E. Cárcamo~Hernández, H.~N. Long, and V.~V. Vien, ``{The first
  $\Delta(27)$ flavor 3-3-1 model with low scale seesaw mechanism},''
  \href{http://dx.doi.org/10.1140/epjc/s10052-018-6284-0}{{\em Eur. Phys. J.}
  {\bfseries C78} no.~10, (2018) 804},
\href{http://arxiv.org/abs/1803.01636}{{\ttfamily arXiv:1803.01636 [hep-ph]}}.

\bibitem{Emmanuel-Costa:2013gia}
D.~Emmanuel-Costa, C.~Simoes, and M.~Tortola, ``{The minimal adjoint-SU(5) x
  $Z_{4}$ GUT model},'' \href{http://dx.doi.org/10.1007/JHEP10(2013)054}{{\em
  JHEP} {\bfseries 10} (2013) 054},
\href{http://arxiv.org/abs/1303.5699}{{\ttfamily arXiv:1303.5699 [hep-ph]}}.

\bibitem{Arbelaez:2015toa}
C.~Arbelaez, A.~E. Carcamo~Hernandez, S.~Kovalenko, and I.~Schmidt, ``{Adjoint
  $SU(5)$ GUT model with $T_{7}$ flavor symmetry},''
  \href{http://dx.doi.org/10.1103/PhysRevD.92.115015}{{\em Phys. Rev.}
  {\bfseries D92} no.~11, (2015) 115015},
\href{http://arxiv.org/abs/1507.03852}{{\ttfamily arXiv:1507.03852 [hep-ph]}}.

\bibitem{CarcamoHernandez:2018aon}
A.~E. Carcamo~Hernandez and S.~F. King, ``{Muon anomalies and the $SU(5)$
  Yukawa relations},''
\href{http://arxiv.org/abs/1803.07367}{{\ttfamily arXiv:1803.07367 [hep-ph]}}.

\bibitem{Hernandez:2014vta}
A.~E. Carcamo~Hernandez, R.~Martinez, and J.~Nisperuza, ``{$S_3$ discrete group
  as a source of the quark mass and mixing pattern in $331$ models},''
  \href{http://dx.doi.org/10.1140/epjc/s10052-015-3278-z}{{\em Eur. Phys. J.}
  {\bfseries C75} no.~2, (2015) 72},
\href{http://arxiv.org/abs/1401.0937}{{\ttfamily arXiv:1401.0937 [hep-ph]}}.

\bibitem{CarcamoHernandez:2017cwi}
A.~E. Carcamo~Hernandez, S.~Kovalenko, H.~N. Long, and I.~Schmidt, ``{A variant
  of 3-3-1 model for the generation of the SM fermion mass and mixing
  pattern},'' \href{http://dx.doi.org/10.1007/JHEP07(2018)144}{{\em JHEP}
  {\bfseries 07} (2018) 144},
\href{http://arxiv.org/abs/1705.09169}{{\ttfamily arXiv:1705.09169 [hep-ph]}}.

\bibitem{CarcamoHernandez:2019cbd}
A.~E. Cárcamo~Hernández, S.~Kovalenko, R.~Pasechnik, and I.~Schmidt,
  ``{Sequentially loop-generated quark and lepton mass hierarchies in an
  extended Inert Higgs Doublet model},''
\href{http://arxiv.org/abs/1901.02764}{{\ttfamily arXiv:1901.02764 [hep-ph]}}.

\bibitem{Hernandez:2015zeh}
A.~E. Cárcamo~Hernández, I.~de~Medeiros~Varzielas, and N.~A. Neill, ``{Novel
  Randall-Sundrum model with $S_{3}$ flavor symmetry},''
  \href{http://dx.doi.org/10.1103/PhysRevD.94.033011}{{\em Phys. Rev.}
  {\bfseries D94} no.~3, (2016) 033011},
\href{http://arxiv.org/abs/1511.07420}{{\ttfamily arXiv:1511.07420 [hep-ph]}}.

\bibitem{Arbelaez:2016mhg}
C.~Arbeláez, A.~E. Cárcamo~Hernández, S.~Kovalenko, and I.~Schmidt,
  ``{Radiative Seesaw-type Mechanism of Fermion Masses and Non-trivial Quark
  Mixing},'' \href{http://dx.doi.org/10.1140/epjc/s10052-017-4948-9}{{\em Eur.
  Phys. J.} {\bfseries C77} no.~6, (2017) 422},
\href{http://arxiv.org/abs/1602.03607}{{\ttfamily arXiv:1602.03607 [hep-ph]}}.

\bibitem{Campos:2014zaa}
M.~D. Campos, A.~E. Carcamo~Hernandez, H.~Pas, and E.~Schumacher, ``{Higgs
  $\rightarrow$ $\mu\tau$ as an indication for $S_4$ flavor symmetry},''
  \href{http://dx.doi.org/10.1103/PhysRevD.91.116011}{{\em Phys. Rev.}
  {\bfseries D91} no.~11, (2015) 116011},
\href{http://arxiv.org/abs/1408.1652}{{\ttfamily arXiv:1408.1652 [hep-ph]}}.

\bibitem{Hernandez:2015tna}
A.~E. Carcamo~Hernandez and R.~Martinez, ``{A predictive 3-3-1 model with $A_4$
  flavor symmetry},''
  \href{http://dx.doi.org/10.1016/j.nuclphysb.2016.02.025}{{\em Nucl. Phys.}
  {\bfseries B905} (2016) 337--358},
\href{http://arxiv.org/abs/1501.05937}{{\ttfamily arXiv:1501.05937 [hep-ph]}}.

\bibitem{CarcamoHernandez:2018vdj}
A.~E. Carcamo~Hernandez, J.~Vignatti, and A.~Zerwekh, ``{A model of strongly
  coupled heavy vector resonances for fermion masses and mixings},''
\href{http://arxiv.org/abs/1807.05321}{{\ttfamily arXiv:1807.05321 [hep-ph]}}.

\bibitem{Chauhan:2018uuy}
G.~Chauhan, P.~S.~B. Dev, R.~N. Mohapatra, and Y.~Zhang, ``{Perturbativity
  constraints on $U(1)_{B-L}$ and left-right models and implications for heavy
  gauge boson searches},''
\href{http://arxiv.org/abs/1811.08789}{{\ttfamily arXiv:1811.08789 [hep-ph]}}.

\bibitem{Grimus:2000vj}
W.~Grimus and L.~Lavoura, ``{The Seesaw mechanism at arbitrary order:
  Disentangling the small scale from the large scale},''
  \href{http://dx.doi.org/10.1088/1126-6708/2000/11/042}{{\em JHEP} {\bfseries
  11} (2000) 042},
\href{http://arxiv.org/abs/hep-ph/0008179}{{\ttfamily arXiv:hep-ph/0008179
  [hep-ph]}}.

\bibitem{Ma:2006km}
E.~Ma, ``{Verifiable radiative seesaw mechanism of neutrino mass and dark
  matter},'' \href{http://dx.doi.org/10.1103/PhysRevD.73.077301}{{\em Phys.
  Rev.} {\bfseries D73} (2006) 077301},
\href{http://arxiv.org/abs/hep-ph/0601225}{{\ttfamily arXiv:hep-ph/0601225
  [hep-ph]}}.

\bibitem{Grimus:2003yn}
W.~Grimus and L.~Lavoura, ``{A Nonstandard CP transformation leading to maximal
  atmospheric neutrino mixing},''
  \href{http://dx.doi.org/10.1016/j.physletb.2003.10.075}{{\em Phys. Lett.}
  {\bfseries B579} (2004) 113--122},
\href{http://arxiv.org/abs/hep-ph/0305309}{{\ttfamily arXiv:hep-ph/0305309
  [hep-ph]}}.

\bibitem{Schechter:1981cv}
J.~Schechter and J.~W.~F. Valle, ``{Neutrino Decay and Spontaneous Violation of
  Lepton Number},'' \href{http://dx.doi.org/10.1103/PhysRevD.25.774}{{\em Phys.
  Rev.} {\bfseries D25} (1982) 774}.

\bibitem{babu:2002dz}
K.~S. Babu, E.~Ma, and J.~W.~F. Valle, ``{Underlying A(4) symmetry for the
  neutrino mass matrix and the quark mixing matrix},''
  \href{http://dx.doi.org/10.1016/S0370-2693(02)03153-2}{{\em Phys. Lett.}
  {\bfseries B552} (2003) 207--213},
\href{http://arxiv.org/abs/hep-ph/0206292}{{\ttfamily arXiv:hep-ph/0206292
  [hep-ph]}}.

\bibitem{Chen:2015siy}
P.~Chen, G.-J. Ding, F.~Gonzalez-Canales, and J.~W.~F. Valle, ``{Generalized
  $\mu-\tau$ reflection symmetry and leptonic CP violation},''
  \href{http://dx.doi.org/10.1016/j.physletb.2015.12.069}{{\em Phys. Lett.}
  {\bfseries B753} (2016) 644--652},
\href{http://arxiv.org/abs/1512.01551}{{\ttfamily arXiv:1512.01551 [hep-ph]}}.

\bibitem{Chen:2016ica}
P.~Chen, G.-J. Ding, F.~Gonzalez-Canales, and J.~W.~F. Valle, ``{Classifying CP
  transformations according to their texture zeros: theory and implications},''
  \href{http://dx.doi.org/10.1103/PhysRevD.94.033002}{{\em Phys. Rev.}
  {\bfseries D94} no.~3, (2016) 033002},
\href{http://arxiv.org/abs/1604.03510}{{\ttfamily arXiv:1604.03510 [hep-ph]}}.

\bibitem{Schechter:1980gr}
J.~Schechter and J.~W.~F. Valle, ``{Neutrino Masses in SU(2) x U(1)
  Theories},''
\href{http://dx.doi.org/10.1103/PhysRevD.22.2227}{{\em Phys. Rev.} {\bfseries
  D22} (1980) 2227}.

\bibitem{Rodejohann:2011vc}
W.~Rodejohann and J.~W.~F. Valle, ``{Symmetrical Parametrizations of the Lepton
  Mixing Matrix},'' \href{http://dx.doi.org/10.1103/PhysRevD.84.073011}{{\em
  Phys. Rev.} {\bfseries D84} (2011) 073011},
\href{http://arxiv.org/abs/1108.3484}{{\ttfamily arXiv:1108.3484 [hep-ph]}}.

\bibitem{Gomez-Izquierdo:2017rxi}
J.~C. Gómez-Izquierdo, ``{Non-minimal flavored ${S}_{3}\otimes {Z}_{2}$
  left–right symmetric model},''
  \href{http://dx.doi.org/10.1140/epjc/s10052-017-5094-0}{{\em Eur. Phys. J.}
  {\bfseries C77} no.~8, (2017) 551},
\href{http://arxiv.org/abs/1701.01747}{{\ttfamily arXiv:1701.01747 [hep-ph]}}.

\bibitem{Garces:2018nar}
E.~A. Garcés, J.~C. Gómez-Izquierdo, and F.~Gonzalez-Canales, ``{Flavored
  non-minimal left–right symmetric model fermion masses and mixings},''
  \href{http://dx.doi.org/10.1140/epjc/s10052-018-6271-5}{{\em Eur. Phys. J.}
  {\bfseries C78} no.~10, (2018) 812},
\href{http://arxiv.org/abs/1807.02727}{{\ttfamily arXiv:1807.02727 [hep-ph]}}.

\bibitem{KamLAND-Zen:2016pfg}
{\bfseries KamLAND-Zen} Collaboration, A.~Gando {\em et~al.}, ``{Search for
  Majorana Neutrinos near the Inverted Mass Hierarchy Region with
  KamLAND-Zen},'' \href{http://dx.doi.org/10.1103/PhysRevLett.117.109903,
  10.1103/PhysRevLett.117.082503}{{\em Phys. Rev. Lett.} {\bfseries 117} no.~8,
  (2016) 082503}, \href{http://arxiv.org/abs/1605.02889}{{\ttfamily
  arXiv:1605.02889 [hep-ex]}}.
[Addendum: Phys. Rev. Lett.117,no.10,109903(2016)].

\bibitem{GERDA:2018}
{\bfseries GERDA Collaboration} Collaboration, M.~Agostini {\em et~al.},
  ``{Improved Limit on Neutrinoless Double-beta decay of 76Ge from GERDA Phase
  II},'' {\em Phys. Rev. Lett.} {\bfseries 120} (2018) 132503.

\bibitem{MAJORANA:2008}
{\bfseries MAJORANA Collaboration} Collaboration, C.~E. Aalseth {\em et~al.},
  ``{Search for Zero-Neutrino Double Beta Decay in $^{76}$Ge with the Majorana
  Demonstrator},''.

\bibitem{CUORE:2018}
{\bfseries CUORE Collaboration} Collaboration, C.~Alduino {\em et~al.},
  ``{First Results from CUORE: A Search for Lepton Number Violation via
  $0\nu\beta\beta$ Decay of $^{130}$Te},'' {\em Phys. Rev. Lett.} {\bfseries
  120} (2018) 132501.

\bibitem{EXO-2018}
{\bfseries EXO-200 Collaboration} Collaboration, J.~Albert {\em et~al.},
  ``{Search for Neutrinoless Double-Beta Decay with the Upgraded EXO-200
  Detector},'' {\em Phys. Rev. Lett.} {\bfseries 120} (2018) 072701.

\bibitem{Arnold:2016bed}
{\bfseries NEMO-3} Collaboration, R.~Arnold {\em et~al.}, ``{Measurement of the
  $2\nu\beta\beta$ Decay Half-Life and Search for the $0\nu\beta\beta$ Decay of
  $^{116}$Cd with the NEMO-3 Detector},''
  \href{http://dx.doi.org/10.1103/PhysRevD.95.012007}{{\em Phys. Rev.}
  {\bfseries D95} no.~1, (2017) 012007},
\href{http://arxiv.org/abs/1610.03226}{{\ttfamily arXiv:1610.03226 [hep-ex]}}.

\bibitem{Alduino:2017pni}
{\bfseries CUORE} Collaboration, C.~Alduino {\em et~al.}, ``{CUORE sensitivity
  to $0\nu \beta \beta $ decay},''
  \href{http://dx.doi.org/10.1140/epjc/s10052-017-5098-9}{{\em Eur. Phys. J.}
  {\bfseries C77} no.~8, (2017) 532},
\href{http://arxiv.org/abs/1705.10816}{{\ttfamily arXiv:1705.10816
  [physics.ins-det]}}.

\bibitem{Albert:2017owj}
{\bfseries EXO} Collaboration, J.~B. Albert {\em et~al.}, ``{Search for
  Neutrinoless Double-Beta Decay with the Upgraded EXO-200 Detector},''
  \href{http://dx.doi.org/10.1103/PhysRevLett.120.072701}{{\em Phys. Rev.
  Lett.} {\bfseries 120} no.~7, (2018) 072701},
\href{http://arxiv.org/abs/1707.08707}{{\ttfamily arXiv:1707.08707 [hep-ex]}}.

\bibitem{Abt:2004yk}
I.~Abt {\em et~al.}, ``{A New $Ge^{76}$ Double Beta Decay Experiment at LNGS:
  Letter of Intent},''
\href{http://arxiv.org/abs/hep-ex/0404039}{{\ttfamily arXiv:hep-ex/0404039
  [hep-ex]}}.

\bibitem{Gilliss:2018lke}
{\bfseries MAJORANA} Collaboration, T.~Gilliss {\em et~al.}, ``{Recent Results
  from the Majorana Demonstrator},''
  \href{http://dx.doi.org/10.1142/S2010194518600492}{{\em Int. J. Mod. Phys.
  Conf. Ser.} {\bfseries 46} (2018) 1860049},
\href{http://arxiv.org/abs/1804.01582}{{\ttfamily arXiv:1804.01582
  [physics.ins-det]}}.

\bibitem{Bora:2012tx}
K.~Bora, ``{Updated values of running quark and lepton masses at GUT scale in
  SM, 2HDM and MSSM},'' {\em Horizon} {\bfseries 2} (2013) ,
\href{http://arxiv.org/abs/1206.5909}{{\ttfamily arXiv:1206.5909 [hep-ph]}}.

\bibitem{Xing:2007fb}
Z.-z. Xing, H.~Zhang, and S.~Zhou, ``{Updated Values of Running Quark and
  Lepton Masses},'' \href{http://dx.doi.org/10.1103/PhysRevD.77.113016}{{\em
  Phys. Rev.} {\bfseries D77} (2008) 113016},
\href{http://arxiv.org/abs/0712.1419}{{\ttfamily arXiv:0712.1419 [hep-ph]}}.

\bibitem{Olive:2016xmw}
{\bfseries Particle Data Group} Collaboration, C.~Patrignani {\em et~al.},
  ``{Review of Particle Physics},''
\href{http://dx.doi.org/10.1088/1674-1137/40/10/100001}{{\em Chin. Phys.}
  {\bfseries C40} no.~10, (2016) 100001}.

\bibitem{Das:2012ii}
S.~Das {\em et~al.}, ``{Heavy Neutrinos and Lepton Flavour Violation in
  Left-Right Symmetric Models at the LHC},'' {\em Phys.Rev.} {\bfseries D86}
  055006, \href{http://arxiv.org/abs/1206.0256}{{\ttfamily arXiv:1206.0256
  [hep-ph]}}.

\bibitem{Deppisch:2013cya}
F.~F. Deppisch, N.~Desai, and J.~W.~F. Valle, ``{Is charged lepton flavour
  violation a high energy phenomenon?},'' {\em Phys.Rev.} {\bfseries D89}
  051302(R), \href{http://arxiv.org/abs/1308.6789}{{\ttfamily arXiv:1308.6789
  [hep-ph]}}.

\end{thebibliography}\endgroup


\providecommand{\href}[2]{#2}\begingroup\raggedright\endgroup
\end{document}